\documentclass{aa}

\usepackage{graphicx}
\usepackage{amsmath,amsfonts,amssymb}
\usepackage{esvect}
\usepackage{txfonts}
\usepackage{color}
\usepackage{natbib}
\usepackage{subcaption}
\usepackage{float}
\usepackage{dblfloatfix}
\usepackage{afterpage}
\usepackage{ifthen}
\usepackage[morefloats=12]{morefloats}
\usepackage{placeins}
\usepackage{multicol}
\usepackage{siunitx}
\DeclareSIUnit\parsec{pc}
\DeclareSIUnit\lightyear{ly}
\DeclareSIUnit\Kcmb{K_{cmb}}
\DeclareSIUnit\year{yr}
\bibpunct{(}{)}{;}{a}{}{,}
\usepackage[switch]{lineno}
\definecolor{linkcolor}{rgb}{0.6,0,0}
\definecolor{citecolor}{rgb}{0,0,0.75}
\definecolor{urlcolor}{rgb}{0.12,0.46,0.7}
\usepackage[breaklinks, colorlinks, urlcolor=urlcolor,
    linkcolor=linkcolor,citecolor=citecolor,pdfencoding=auto]{hyperref}
\hypersetup{linktocpage}

\def\setsymbol#1#2{\expandafter\def\csname #1\endcsname{#2}}
\def\getsymbol#1{\csname #1\endcsname}

\newbox\tablebox    \newdimen\tablewidth
\def\leaderfil{\leaders\hbox to 5pt{\hss.\hss}\hfil}
\def\tablenote#1 #2\par{\begingroup \parindent=0.8em
    \abovedisplayshortskip=0pt\belowdisplayshortskip=0pt
    \noindent
    $$\hss\vbox{\hsize\tablewidth \hangindent=\parindent \hangafter=1 \noindent
    \hbox to \parindent{$^#1$\hss}\strut#2\strut\par}\hss$$
    \endgroup}

\def\L2{\ifmmode L_2\else $L_2$\fi}

\def\DeltaT{\ifmmode \Delta T\else $\Delta T$\fi}
\def\deltat{\ifmmode \Delta t\else $\Delta t$\fi}
\def\fknee{\ifmmode f_{\rm knee}\else $f_{\rm knee}$\fi}
\def\Fmax{\ifmmode F_{\rm max}\else $F_{\rm max}$\fi}
\def\solar{\ifmmode{\rm M}_{\mathord\odot}\else${\rm M}_{\mathord\odot}$\fi}
\def\Msolar{\ifmmode{\rm M}_{\mathord\odot}\else${\rm M}_{\mathord\odot}$\fi}
\def\Lsolar{\ifmmode{\rm L}_{\mathord\odot}\else${\rm L}_{\mathord\odot}$\fi}
\def\inv{\ifmmode^{-1}\else$^{-1}$\fi}
\def\mo{\ifmmode^{-1}\else$^{-1}$\fi}
\def\sup#1{\ifmmode ^{\rm #1}\else $^{\rm #1}$\fi}
\def\expo#1{\ifmmode \times 10^{#1}\else $\times 10^{#1}$\fi}
\def\,{\thinspace}
\def\lsim{\mathrel{\raise .4ex\hbox{\rlap{$<$}\lower 1.2ex\hbox{$\sim$}}}}
\def\gsim{\mathrel{\raise .4ex\hbox{\rlap{$>$}\lower 1.2ex\hbox{$\sim$}}}}

\def\simprop{\mathrel{\raise .4ex\hbox{\rlap{$\propto$}\lower 1.2ex\hbox{$\sim$}}}}
\def\deg{\ifmmode^\circ\else$^\circ$\fi}
\def\pdeg{\ifmmode $\setbox0=\hbox{$^{\circ}$}\rlap{\hskip.11\wd0 .}$^{\circ}
          \else \setbox0=\hbox{$^{\circ}$}\rlap{\hskip.11\wd0 .}$^{\circ}$\fi}
\def\arcs{\ifmmode {^{\scriptstyle\prime\prime}}
          \else $^{\scriptstyle\prime\prime}$\fi}
\def\arcm{\ifmmode {^{\scriptstyle\prime}}
          \else $^{\scriptstyle\prime}$\fi}
\newdimen\sa  \newdimen\sb
\def\parcs{\sa=.07em \sb=.03em
     \ifmmode \hbox{\rlap{.}}^{\scriptstyle\prime\kern -\sb\prime}\hbox{\kern -\sa}
     \else \rlap{.}$^{\scriptstyle\prime\kern -\sb\prime}$\kern -\sa\fi}
\def\parcm{\sa=.08em \sb=.03em
     \ifmmode \hbox{\rlap{.}\kern\sa}^{\scriptstyle\prime}\hbox{\kern-\sb}
     \else \rlap{.}\kern\sa$^{\scriptstyle\prime}$\kern-\sb\fi}
\def\ra[#1 #2 #3.#4]{#1\sup{h}#2\sup{m}#3\sup{s}\llap.#4}
\def\dec[#1 #2 #3.#4]{#1\deg#2\arcm#3\arcs\llap.#4}
\def\deco[#1 #2 #3]{#1\deg#2\arcm#3\arcs}
\def\rra[#1 #2]{#1\sup{h}#2\sup{m}}

\def\dots{\relax\ifmmode \ldots\else $\ldots$\fi}
\def\WHzsr{\ifmmode $W\,Hz\mo\,sr\mo$\else W\,Hz\mo\,sr\mo\fi}
\def\mHz{\ifmmode $\,mHz$\else \,mHz\fi}
\def\GHz{\ifmmode $\,GHz$\else \,GHz\fi}
\def\mKs{\ifmmode $\,mK\,s$^{1/2}\else \,mK\,s$^{1/2}$\fi}
\def\muKs{\ifmmode \,\mu$K\,s$^{1/2}\else \,$\mu$K\,s$^{1/2}$\fi}
\def\muKRJs{\ifmmode \,\mu$K$_{\rm RJ}$\,s$^{1/2}\else \,$\mu$K$_{\rm RJ}$\,s$^{1/2}$\fi}
\def\muKHz{\ifmmode \,\mu$K\,Hz$^{-1/2}\else \,$\mu$K\,Hz$^{-1/2}$\fi}
\def\MJysr{\ifmmode \,$MJy\,sr\mo$\else \,MJy\,sr\mo\fi}
\def\MJysrmK{\ifmmode \,$MJy\,sr\mo$\,mK$_{\rm CMB}\mo\else \,MJy\,sr\mo\,mK$_{\rm CMB}\mo$\fi}
\def\microns{\ifmmode \,\mu$m$\else \,$\mu$m\fi}

\def\muK{\ifmmode \,\mu$K$\else \,$\mu$\hbox{K}\fi}
\def\microK{\ifmmode \,\mu$K$\else \,$\mu$\hbox{K}\fi}
\def\muW{\ifmmode \,\mu$W$\else \,$\mu$\hbox{W}\fi}
\def\kms{\ifmmode $\,km\,s$^{-1}\else \,km\,s$^{-1}$\fi}
\def\kmsMpc{\ifmmode $\,\kms\,Mpc\mo$\else \,\kms\,Mpc\mo\fi}
\providecommand{\sorthelp}[1]{}

\def\fsky{f_{\textrm{sky}}}
\def\WMAP{WMAP}

\def\LCDM{$\Lambda$CDM}

\newcommand{\npipe}[0]{\texttt{NPIPE}}

\newcommand{\oslo}[0]{1}
\newcommand{\planck}[0]{\textit{Planck}}

\let\vec\vv

\def\inv{^{-1}}

\begin{document}

\title{Frequency-dependent constraints on cosmic birefringence from the LFI and HFI \planck\ Data Release 4}

\author{\small
J.~R.~Eskilt\inst{\oslo}\thanks{Corresponding author: J.~R.~Eskilt; \url{j.r.eskilt@astro.uio.no}}
}
\institute{\small
Institute of Theoretical Astrophysics, University of Oslo, P.O. Box 1029 Blindern, N-0315 Oslo, Norway\goodbreak
}
\authorrunning{Eskilt}
\titlerunning{Frequency-Dependent Cosmic Birefringence}

\abstract{We present new constraints on the frequency dependence of the cosmic birefringence angle from the \planck\ Data Release 4 polarization maps. An axion field coupled to electromagnetism predicts a nearly frequency-independent birefringence angle, $\beta_\nu = \beta$, while Faraday rotation from local magnetic fields and Lorentz violating theories predict a cosmic birefringence angle that is proportional to the frequency, $\nu$, to the power of some integer $n$, $\beta_\nu \propto \nu^n$. In this work, we first sampled $\beta_\nu$ individually for each polarized HFI frequency band in addition to the 70\,GHz channel from the LFI. We also constrained a power law formula for the birefringence angle, ${\beta_\nu=\beta_0(\nu/\nu_0)^n}$, with $\nu_0 = 150$\,GHz. For a nearly full-sky measurement, $\fsky=0.93$, we find $\beta_0 = 0.26^{\circ}\pm0.11^\circ$ $(68\% \textrm{ C.L.})$ and ${n=-0.45^{+0.61}_{-0.82}}$ when we ignore the intrinsic $EB$ correlations of the polarized foreground emission, and $\beta_0 = 0.33^\circ \pm 0.12^\circ$ and ${n=-0.37^{+0.49}_{-0.64}}$ when we use a filamentary dust model for the foreground $EB$. Next, we used all the polarized \planck\ maps, including the 30 and 44\,GHz frequency bands. These bands have a negligible foreground contribution from polarized dust emission and we thus treated them separately. Without any modeling of the intrinsic $EB$ of the foreground, we generally find that the inclusion of the 30 and 44\,GHz frequency bands raises the measured values of $\beta_\nu$ and tightens $n$. At nearly full-sky, we measure $\beta_0=0.29^{\circ+0.10^\circ}_{\phantom{\circ}-0.11^\circ}$ and $n=-0.35^{+0.48}_{-0.47}$. Assuming no frequency dependence, we measure $\beta=0.33^\circ \pm 0.10^\circ$. If our measurements have effectively mitigated the $EB$ of the foreground, our constraints are consistent with a mostly frequency-independent signal of cosmic birefringence.
}


\maketitle


\section{Introduction}
\label{sec:introduction}

Parity violation has so far only been observed in the weak interaction, but there are extensions of the standard model that introduce more parity-violating effects. One popular example is a pseudo-scalar axion-like field $\phi$ which couples to the electromagnetic tensor and can cause parity-violating physics for electromagnetic waves \citep{Turner:1987bw, Carroll:1989vb, Harari:1992ea}. The Chern-Simons term couples the field $\phi$ to the electromagnetic tensor $F_{\mu \nu}$ of the form $\mathcal{L} \supset \frac{1}{4}g_{\phi \gamma} \phi F_{\mu \nu} \Tilde{F}^{\mu \nu}$, where $\Tilde{F}_{\mu \nu}$ is the dual tensor of $F_{\mu \nu}$ and $g_{\phi \gamma}$ is the coupling constant (see the review by \citealp{Marsh:2015xka}). It is possible to show that this term changes the phase velocity of the left- and right-handed circular polarization of photons, which effectively causes a rotation of the linear polarization of electromagnetic waves by an angle $\beta$. This can be detected in the polarization of the cosmic microwave background (CMB). The standard model of cosmology, \LCDM, predicts no intrinsic $EB$ correlation of the CMB.

A cosmic birefringence angle $\beta$ would rotate the intrinsic $EE$ power spectrum, $C^{EE}_\ell$, and $BB$ power spectrum, $C^{BB}_\ell$, into an observed $EB$ power spectrum, $C^{EB\textrm{,o}}_\ell = \frac{\sin(4\beta)}{2}\left( C^{EE}_\ell - C^{BB}_\ell\right)$, where ``o'' denotes the observed value and $\ell$ is the multipole moment. However, the problem of measuring $\beta$ this way is two-fold: there is polarized Galactic emission that conceals the cosmic signal and no instruments are perfectly calibrated. The polarization miscalibration angle of an instrument, $\alpha$, is perfectly degenerate with the cosmic birefringence angle, $\beta$, when solely the CMB is analyzed. A lack of strong priors on $\alpha$ for an instrument will necessarily give a large degree of uncertainty on $\beta$. This is seen in the original \planck\ 2015 analysis, where the authors found $\beta = 0.29 \pm 0.05 (\textrm{stat.}) \pm 0.28 (\textrm{syst.})$ \citep{Aghanim:2016fhp}. Here, the latter systematic uncertainty is the miscalibration uncertainty expected from the calibration before launch. The other problem is that the Galactic foreground emission is polarized; without a proper foreground removal, the polarized foreground emission will inevitably contaminate the measurements.

A novel method to mitigate these two problems was developed by \cite{Minami:2019ruj} and \cite{MinamiKomatsu:2020}. The assumption is that the mechanism that can cause cosmic birefringence mostly affects the CMB photons. This could either be because the effect was the strongest shortly after the last-scattering surface or because the effect is stronger over longer distances. This means that cosmic birefringence has a negligible effect on the Galactic foreground emission and that the linearly polarized CMB emission is rotated by $\alpha + \beta$, while the polarized foreground emission is only rotated by $\alpha$. This method allows for the degeneracy between $\alpha$ and $\beta$ to be broken, but the intrinsic foreground $EB$ power spectrum needed to be specified. Assuming that the intrinsic $EB$ correlation of the foreground is zero, \cite{Minami:2020odp} used the high-frequency instrument (HFI) polarized maps from the \planck\ public release 3 (PR3) to find a measurement of $\beta = 0.34^\circ \pm 0.14^\circ$ (68\%~C.L.), which corresponds to the statistical significance of $2.4\sigma$. They emphasized the possibility that the intrinsic foreground $EB$ can create an effective angle $\gamma$ which skews the measurements $\beta \rightarrow \beta - \gamma$ and $\alpha \rightarrow \alpha + \gamma$; this leaves the sum $\alpha + \beta$ invariant. The authors pointed out that the positive $TB$ and $TE$ inferred from the \planck\ data suggest $EB>0$ \citep{Huffenberger:2019mjx}; hence, $\gamma>0$, which would make the measured $\beta$ a lower bound.

An astrophysical model for the intrinsic $EB$ of dust was described by \cite{Clark:2021kze}. Their work showed evidence for the applied Galactic mask to greatly affect the magnitude and sign of the $EB$ power spectra of the polarized foreground emission. With a Galactic mask of sky fraction $\fsky \approx 0.7$, they found a robustly positive $C^{EB, \text{dust}}_\ell$ for $\ell \lesssim 500$ using their filamentary dust model, which would also suggest a lower bound of $\beta$ if one ignores the intrinsic dust $EB$. On the other hand, a smaller Galactic mask gave $EB$ correlations with a fluctuating sign over multipoles $\ell$. This mask dependence was confirmed by \cite{2022arXiv220107682D} where a declining measurement of $\beta$ was found when ignoring the intrinsic $EB$ power spectra of the foreground using polarized data from the \planck\ public release 4 (PR4) \citep{Akrami:2020bpw}, often called the ``\npipe'' release.

In \citet{Minami:2020odp} and \citet{2022arXiv220107682D}, the cosmic birefringence angle $\beta$ was assumed to be independent of the photon frequency $\nu$. In this paper, we explore a possible frequency dependence. This is not the first time a frequency-dependent cosmic birefringence angle has been considered in the literature. \citet{Gubitosi:2014cua} used polarization data of \WMAP{7}, BOOMERanG, QUAD, and BICEP to look for any hints of a frequency-dependent $\beta$. Other constraints have come from looking at Lorentz violating origins, such as \cite{Kahniashvili:2008va} using \WMAP\ data. \citet{Galaverni:2014gca} also constrained several models of cosmic birefringence using CMB and other astrophysical data at a much wider frequency range. No detection of a frequency-dependent $\beta$ has been found, but this work generalizes the methods of \cite{2022arXiv220107682D} and \cite{Minami:2020odp} to search for such a frequency dependence. These authors assumed that $\beta$ was constant for all frequencies, which gave them a hint of a non-zero $\beta$. Motivated by their result, the goal of this work is to learn more about the frequency dependence of that signal. Also, for the first time, we include the low-frequency instrument (LFI) polarized data to put constraints on the frequency dependence of the signal. All uncertainties quoted in this paper are at a $68\%$ confidence level.

\section{Theory}

In this section, the focus is mainly on an axion-like field that couples to electromagnetism. We also go on to show how we measure the cosmic birefringence signal to be consistent with frequency independence, which we go on to show is consistent with the predictions of an axion-like field. Therefore, we do not go into the details of Faraday rotation or any other models that would cause a frequency-dependent signal.

In this paper, we assume that only the linearly polarized CMB emission has been rotated by the cosmic birefringence angle, $\beta$, while the polarized foreground emission from the Milky Way has not been rotated. We are, therefore, looking for a potential mechanism that is either substantial on large length scales so that its effect on the polarized foreground emission is negligible or it is, instead, an effect that was only present around the time of recombination or shortly after.

An axion-like field $\phi$ that couples to electromagnetism can be the cause of cosmic birefringence. We make the assumption that the field is homogeneous and only varies over time. The dispersion relation for electromagnetic waves becomes \citep{Carroll:1989vb, Harari:1992ea}
\begin{equation}
    \omega^2_{\pm} = k^2 \pm k g_{\phi \gamma} \dot{\phi},
\end{equation}
where $\omega$ is the angular frequency, $k$ is the wavenumber, and $g_{\phi \gamma}$ is the coupling constant; in addition, $+$ is right-handed circular polarization and $-$ is left-handed. Performing a series expansion of $\omega_{\pm}$ in small $g_{\phi \gamma}\dot{\phi}$ gives:
\begin{equation}
    \omega_{\pm} = k \pm \frac{g_{\phi \gamma} \dot{\phi}}{2} - \frac{\left(g_{\phi \gamma} \dot{\phi}\right)^2}{8k} \pm \frac{\left(g_{\phi \gamma} \dot{\phi}\right)^3}{16k^2} + \mathcal{O}\left(\left(g_{\phi \gamma} \dot{\phi} \right)^4/ k^3\right).
\end{equation}
Our convention for the cosmic birefringence angle $\beta$ is that it rotates the linear polarization clockwise on the sky. We then get
\begin{align}
    \nonumber
    \beta &= -\int dt \, \frac{\omega_+ - \omega_-}{2}\\
    \label{eq:beta_pert}
    &= -\int dt \, \left[\frac{g_{\phi \gamma} \dot{\phi}}{2} + \frac{\left(g_{\phi \gamma} \dot{\phi}\right)^3}{16k^2}+ \mathcal{O}\left(\left(g_{\phi \gamma} \dot{\phi} \right)^5/ k^4\right)\right].
\end{align}
The first term is independent of the wavenumber. Neglecting any electromagnetic contribution, the equation of motion for $\phi$ is given by the Klein-Gordon equation \citep{Marsh:2015xka},
\begin{equation}
    \label{eq:klein-gordon}
    \Ddot{\phi} + 3H\dot{\phi} + m^2\phi=0,
\end{equation}
where $m$ is the mass of the axion-like particle, while $H$ is the Hubble parameter. There is only a finite mass interval of $m$ where we can detect an axion-like particle with our method. If the mass is smaller than the Hubble expansion today, $H_0 \approx 10^{-33}$~eV, then the field is generally too slowly varying by the Hubble friction to be detected. We note that a large coupling constant, $g_{\phi\gamma}$, can make fields with smaller masses detectable \citep{Fujita:2020ecn}. On the other hand, to get a significant rotation from $\beta$, we need the field to start oscillating after decoupling which yields $m \lesssim 10^{-28}\, \textrm{eV}$ \citep{Marsh:2015xka, Arvanitaki:2009fg}. Fields that oscillate before the decoupling quickly get suppressed since the era of decoupling was not instantaneous, meaning that not all polarized waves were rotated equally. The method behind this analysis is therefore able to probe ultra-light axion-like fields with a mass range of $10^{-33}\, \textrm{eV} \lesssim m \lesssim 10^{-28}\, \textrm{eV}$ and possibly even lower, depending on the amplitude of $g_{\phi \gamma}$.

If the mass is much larger than $H_0$, the axion-like field has decayed, $\phi(t_0) \approx 0$, leading the cosmic birefringence angle to be $\beta = g_{\phi \gamma} \phi_{\text{rec}}/2$, where $\phi_{\text{rec}}$ is the value of the field at recombination that is expected to have a negligible difference to its primordial value. A detection of cosmic birefringence in polarized CMB data can therefore constrain an axion-like field in the $(g_{\phi \gamma}, \phi_{\text{rec}})$-plane \citep{Marsh:2015xka}.

We go on to derive a crude upper bound on the first frequency-dependent term of $\beta$ in Eq. \eqref{eq:beta_pert} to show that the frequency dependence of $\beta$ cannot be detected by the sensitivity level of \planck. We assume that the Hubble parameter is constant $H=H_{\text{rec}} \approx 10^{-28}\, \textrm{eV}$, and $\phi$ is critically damped, namely, $m=3H/2$. The field is then $\phi = \phi_{\text{rec}} e^{-3H t/2}$, where we take $t=0$ at the time of recombination and approximate today to be at infinite time $t \rightarrow \infty$. The contribution from the second term in Eq.~\eqref{eq:beta_pert} is:
\begin{equation}
    \left|\int dt \,  \frac{\left(g_{\phi \gamma} \dot{\phi}\right)^3}{16k^2} \right| = \frac{3 |g_{\phi \gamma}\phi_{\text{rec}}|^3}{256\pi^2} \left(\frac{H_{\text{rec}}}{(1+z)\nu} \right)^2 = \left(1.7 \cdot 10^{-60}\right)^\circ,
\end{equation}
where the frequency is ${\nu =k/2\pi}$, and we used the lowest \planck\ frequency band that we consider in this work, namely, $\nu = 30$\,GHz; also, we assume that the redshift at recombination is $z\approx 1100$, so that the initial frequency at recombination is $\nu_i = (1+z)\nu$. We also use the result of \cite{Minami:2020odp} that $\beta = -\int dt \, \frac{g_{\phi \gamma} \dot{\phi}}{2} = \frac{g_{\phi \gamma}\phi_{\text{rec}}}{2} \approx 0.3^\circ$. The Hubble parameter, $H$, is certainly not constant and it will decrease over time, causing the field $\phi$ to become underdamped. In fact, Eq. \eqref{eq:klein-gordon} can be solved exactly if we assume the expansion parameter $a \propto t^n$ for some value of $n$. The resulting equation takes the form of Bessel functions \citep{Marsh:2015xka}. The point is however that changing the mass $m$ and allowing $H$ to vary with time will not magnify the first frequency-dependent term by around $60$ orders of magnitude. It is, therefore, safe to assume that \planck's sensitivity level will not pick up any potential frequency dependence of $\beta$ caused by an ultra-light axion-like field coupled to electromagnetism.

There are, however, other mechanisms that could cause cosmic birefringence that we mention here in brief \citep{2013JCAP...02..020G}. If the observed isotropic cosmic birefringence was caused by Faraday rotation originating from primordial magnetic fields \citep{Subramanian:2015lua}, we would expect $\beta_\nu \propto \nu^{-2}$. In the case where the birefringence angle is linear in frequency, $\beta_\nu \propto \nu$, we could be observing Lorentz violating electrodynamics \citep{Shore:2004sh}. Such a measurement could be translated into a test of Lorentz violation \citep{Kahniashvili:2008va}. We also consider the possibility of $\beta_\nu \propto \nu^2$, which has been predicted from quantum gravity theories to potentially modify the dispersion relation of photons \citep{Myers:2003fd}. We are therefore able to learn a lot about the origin of the measured cosmic birefringence result of \cite{2022arXiv220107682D} by quantifying the frequency dependence of the signal.

\section{Method}
\label{sec:method}

Our analysis method follows that of \cite{Minami:2019ruj} and \cite{MinamiKomatsu:2020}. We start with the main assumption that the effect of cosmic birefringence, $\beta$, only substantially affects the CMB photons. However, an instrumental polarization miscalibration angle alters all linearly polarized plane waves, which we characterize by an instrument-dependent angle $\alpha$. For a single detector, we can therefore write the spherical harmonics coefficients for the $E$- and $B$-modes \citep{Seljak:1996gy, Kamionkowski:1996zd} as:
\begin{align}
    \nonumber
    E^o_{\ell m} = &\cos(2\alpha)E^{\text{fg}}_{\ell m} - \sin(2\alpha)B^{\text{fg}}_{\ell m}\\
    +&\cos(2\alpha+2\beta)E^{\text{CMB}}_{\ell m} - \sin(2\alpha+2\beta)B^{\text{CMB}}_{\ell m} + E^{\text{noise}}_{\ell m},\\
    \nonumber
     B^o_{\ell m} = &\cos(2\alpha)B^{\text{fg}}_{\ell m} + \sin(2\alpha)E^{\text{fg}}_{\ell m}\\
     + &\cos(2\alpha+2\beta)B^{\text{CMB}}_{\ell m} + \sin(2\alpha+2\beta)E^{\text{CMB}}_{\ell m}+ B^{\text{noise}}_{\ell m}.
\end{align}
Here, ``fg'' denotes the foreground component. We define the power spectra as $C_\ell^{XY} = \frac{1}{2\ell + 1}\sum_m X_{\ell m}Y^*_{\ell m}$. To account for the effect of noise, we also define the ensemble average of the power spectra as $\langle C^{XY}_\ell \rangle  = \delta_{m m} \delta_{\ell \ell'}\langle X_{\ell m}Y^*_{\ell' m'} \rangle$. Taking the equations above, we can derive the following equation \citep{Minami:2019ruj}:
\begin{align}
    \nonumber
    \langle C_\ell^{EB, o}\rangle &= \frac{\tan(4\alpha)}{2}\left(\langle C_\ell^{EE, o}\rangle - \langle C_\ell^{BB, o}\rangle \right)\\
    \nonumber
    &+ \frac{\sin(4\beta)}{2\cos(4\alpha)}\left(\langle C_\ell^{EE, \text{CMB}}\rangle - \langle C_\ell^{BB, \text{CMB}}\rangle \right)\\
    \label{eq:singleinstrumentbeta}
    &+ \frac{1}{\cos(4\alpha)}\langle C_\ell^{EB, \text{fg}}\rangle  + \frac{\cos(4\beta)}{\cos(4\alpha)}\langle C_\ell^{EB, \text{CMB}}\rangle,
\end{align}
where ``o'' denotes the observed value. We have thus been able to derive an equation that does not explicitly depend on the $E$- and $B$-modes of the polarized foregrounds, $\langle C_\ell^{EE, \text{fg}}\rangle $ and $\langle C_\ell^{BB, \text{fg}}\rangle $. And we can use the foreground to our advantage to break the degeneracy between the miscalibration angle $\alpha$ and the cosmic birefringence angle $\beta$. Unfortunately, we are left with the intrinsic $EB$-mode of the polarized foreground emission in our equation. For lack of an astrophysical model, this was set to zero in the initial analysis of \planck\ Data Release 3 in \cite{Minami:2020odp}, but they gave a discussion of how a non-zero $EB$ would influence the analysis \citep{Minami:2019ruj, MinamiKomatsu:2020}. We later show how the filamentary dust model of \cite{Clark:2021kze} can create non-zero $EB$ correlations of dust and how to implement an ansatz motivated by their model into our equations.

The last term in Eq. \eqref{eq:singleinstrumentbeta} can be assumed to be zero since $\Lambda$CDM does not predict any parity-violating correlations at the last scattering surface. However, chiral gravitational waves \citep{Thorne:2017jft} would predict intrinsic $EB$ correlations of the CMB emissions. Thus, we could, in principle, use Eq. \eqref{eq:singleinstrumentbeta} to probe this term, but for the purpose of this work, we set $\langle C^{EB, \textrm{CMB}}_\ell \rangle=0$.

Before applying this to real data, we need to generalize the method to incorporate multiple frequency bands \citep{MinamiKomatsu:2020}. This work is based on the \npipe\ pipeline which splits the detectors of each of the HFI frequency bands and the 70\,GHz band into two groups, namely the $A$ and $B$ split. Therefore, for each of these frequency bands, we have two maps of the sky and two miscalibration angles per band, which act as the average over the detectors for a given split. We also analyze the frequency dependence of a possible cosmic birefringence angle $\beta_\nu$. Thus, we denote $i$ as a specific frequency band and data split for the miscalibration angle $\alpha_i$, but for $\beta_i$, it only denotes the frequency of a given band. Following \cite{MinamiKomatsu:2020} with the exception of having a frequency-dependent $\beta_i$, we now write observed spherical harmonics coefficients for polarization. For completeness, we derive the equations based on the assumption that we have access to the $EB$ power spectrum of the polarized foreground emission. Assuming that the CMB has no intrinsic $EB$ correlations, we get
\begin{align}
    \nonumber
    \begin{bmatrix}
      \langle C_\ell^{E_iE_j, o}\rangle\\
        \langle C_\ell^{B_iB_j, o}\rangle
    \end{bmatrix} = \mathbf{R}(\alpha_i, \alpha_j) \begin{bmatrix}
        \langle C_\ell^{E_iE_j, \text{fg}}\rangle\\
          \langle C_\ell^{B_iB_j, \text{fg}}\rangle
    \end{bmatrix}+\mathbf{D}(\alpha_i, \alpha_j)\begin{bmatrix}
        \langle C_\ell^{E_iB_j, \text{fg}}\rangle\\
          \langle C_\ell^{B_iE_j, \text{fg}}\rangle
    \end{bmatrix}&\\
    \label{eq:c_ee_c_bb}
    +\mathbf{R}(\alpha_i+\beta_i, \alpha_j+\beta_j) \begin{bmatrix}
        \langle C_\ell^{E_iE_j, \text{CMB}}\rangle\\
          \langle C_\ell^{B_iB_j, \text{CMB}}\rangle
    \end{bmatrix}+\delta_{i,j} \begin{bmatrix}
    \langle C_\ell^{E_iE_j, \text{noise}}\rangle\\
      \langle C_\ell^{B_iB_j, \text{noise}}\rangle
    \end{bmatrix}&,
\end{align}
\begin{align}
        \nonumber
      \langle C_\ell^{E_iB_j, o}\rangle &= R^T(\alpha_i, \alpha_j) \begin{bmatrix}
        \langle C_\ell^{E_iE_j, \text{fg}}\rangle\\
          \langle C_\ell^{B_iB_j, \text{fg}}\rangle
    \end{bmatrix}+D^T(\alpha_i, \alpha_j)\begin{bmatrix}
        \langle C_\ell^{E_iB_j, \text{fg}}\rangle\\
          \langle C_\ell^{B_iE_j, \text{fg}}\rangle
    \end{bmatrix}\\
    \label{eq:c_eb}
    &+R^T(\alpha_i+\beta_i, \alpha_j+\beta_j) \begin{bmatrix}
        \langle C_\ell^{E_iE_j, \text{CMB}}\rangle\\
          \langle C_\ell^{B_iB_j, \text{CMB}}\rangle
    \end{bmatrix}.
\end{align}
Here,
\begin{align}
    \mathbf{R}(\theta_i, \theta_j) &= \begin{bmatrix}
        \cos(2\theta_i)\cos(2\theta_j) & \sin(2\theta_i)\sin(2\theta_j) \\
        \sin(2\theta_i)\sin(2\theta_j) & \cos(2\theta_i)\cos(2\theta_j)
    \end{bmatrix},\\
    \mathbf{D}(\theta_i, \theta_j) &= \begin{bmatrix}
      -\cos(2\theta_i)\sin(2\theta_j) & -\sin(2\theta_i)\cos(2\theta_j) \\
        \sin(2\theta_i)\cos(2\theta_j) & \cos(2\theta_i)\sin(2\theta_j)
    \end{bmatrix},\\
    R(\theta_i, \theta_j) &= \begin{bmatrix}
        \cos(2\theta_i)\sin(2\theta_j)\\
        -\sin(2\theta_i)\cos(2\theta_j)
    \end{bmatrix},\\
    D(\theta_i, \theta_j) &= \begin{bmatrix}
        \cos(2\theta_i)\cos(2\theta_j)\\
        -\sin(2\theta_i)\sin(2\theta_j)
    \end{bmatrix}.
\end{align}
We are not working with auto-spectra, so we can discard the noise term in Eq. \eqref{eq:c_ee_c_bb}. Combining Eqs. \eqref{eq:c_ee_c_bb} and \eqref{eq:c_eb}, we can eliminate $\begin{bmatrix}
        \langle C_\ell^{E_iE_j, \text{fg}}\rangle\\
          \langle C_\ell^{B_iB_j, \text{fg}}\rangle
    \end{bmatrix}$ explicitly to end up with one equation,
\begin{align}
\nonumber
&\langle C_\ell^{E_iB_j, o}\rangle = R^T(\alpha_i, \alpha_j)\textbf{R}^{-1}(\alpha_i, \alpha_j) \begin{bmatrix}
      \langle C_\ell^{E_iE_j, o}\rangle\\
        \langle C_\ell^{B_iB_j, o}\rangle
    \end{bmatrix}+\\
    \nonumber
    &\bigg[R^T(\alpha_i+\beta_i,\alpha_j+\beta_j)-R^T(\alpha_i,\alpha_j)\textbf{R}^{-1}(\alpha_i, \alpha_j)\textbf{R}(\alpha_i+\beta_i, \alpha_j+\beta_j) \bigg]\\
    \label{eq:full_equation}
    &\cdot\begin{bmatrix}
        \langle C_\ell^{E_iE_j, \text{CMB}}\rangle\\
          \langle C_\ell^{B_iB_j, \text{CMB}}\rangle
    \end{bmatrix}
    +Z^T(\alpha_i, \alpha_j)\begin{bmatrix}
        \langle C_\ell^{E_iB_j, \text{fg}}\rangle\\
          \langle C_\ell^{B_iE_j, \text{fg}}\rangle
    \end{bmatrix},
\end{align}
where
\begin{align}
    \nonumber
    Z^T(\alpha_i, \alpha_j) &= D^T(\alpha_i, \alpha_j) -R^T(\alpha_i, \alpha_j) \textbf{R}^{-1}(\alpha_i, \alpha_j)\textbf{D}(\alpha_i, \alpha_j)\\
    &= \frac{2\begin{bmatrix}
        \cos(2\alpha_i)\cos(2\alpha_j), \sin(2\alpha_i)\sin(2\alpha_j)
    \end{bmatrix}}{\cos(4\alpha_i) + \cos(4\alpha_j)}. 
\end{align}

We now switch from the ensemble average spectra, $\langle C_\ell \rangle$, to the estimated power spectra, $C_\ell$. Excluding auto-spectra, we can group all combinations of $(i, j)$ into a vector of observed spectra $\vec{C}^o_\ell = \begin{bmatrix}
  C^{E_i E_j, o}_\ell, C^{B_i B_j, o}_\ell, C^{E_i B_j, o}_\ell
\end{bmatrix}^T$ and vectors of beam-smoothed theoretical $\Lambda \text{CDM}$ spectra $\vec{C}^{\Lambda \text{CDM}}_\ell = \begin{bmatrix}
  C^{E_i E_j, \Lambda \text{CDM}}_\ell, C^{B_i B_j, \Lambda \text{CDM}}_\ell
\end{bmatrix}^T$. If we wish to include the intrinsic $EB$-modes of the foreground, we can make vectors of the polarized foreground $EB$-spectra $\vec{C}^{\text{fg}}_\ell = \begin{bmatrix}
  C^{E_iB_j, \text{fg}}_\ell,C^{B_iE_j, \text{fg}}_\ell
\end{bmatrix}^T$. As an example, we later use ten maps which gives us that the length of $\vec{C}^o_\ell$ is $3 \cdot (10\cdot10 - 10) = 270$, where we excluded the auto spectra, while $\vec{C}^{\Lambda \text{CDM}}_\ell$ and $\vec{C}^{\text{fg}}_\ell$ have $2\cdot(10\cdot10 - 10) = 180$ elements for each multipole $\ell$. To solve this equation for all combinations of input maps, we group them into block diagonal matrices:
\begin{align}
    \textbf{A}_{ij} = &\begin{bmatrix}- R^T(\alpha_i, \alpha_j)\textbf{R}^{-1}(\alpha_i, \alpha_j), 1\end{bmatrix},\\
    \nonumber
    \textbf{B}_{ij} = &\bigg[R^T(\alpha_i+\beta_i,\alpha_j+\beta_j)-R^T(\alpha_i,\alpha_j)
    \textbf{R}^{-1}(\alpha_i, \alpha_j)\\
    &\cdot\textbf{R}(\alpha_i+\beta_i, \alpha_j+\beta_j)\bigg],\\
    \textbf{Z}_{ij} = &Z^T(\alpha_i,\alpha_j),
\end{align}
so that we are solving the equation
\begin{equation}
    \vec{v}^T_{\ell} \equiv \textbf{A}\vec{C}^o_\ell - \textbf{B}\vec{C}^{\Lambda \text{CDM}}_\ell - \textbf{Z}\vec{C}^{\text{fg}}_\ell = 0.
\end{equation}
The covariance matrix for this equation is
\begin{align}
    \nonumber
    \textbf{M}_\ell &= \textbf{A}\text{Cov}(\vec{C}^o_\ell, \vec{C}^o_\ell)\textbf{A}^T + \textbf{B}\text{Cov}(\vec{C}^{\Lambda \text{CDM}}_\ell, \vec{C}^{\Lambda \text{CDM}}_\ell)\textbf{B}^T\\
    \nonumber
    &+ \textbf{Z}\text{Cov}(\vec{C}^{\text{fg}}_\ell, \vec{C}^{\text{fg}}_\ell)\textbf{Z}^T\\
    \nonumber
    &-\textbf{A}\text{Cov}(\vec{C}^o_\ell, \vec{C}^{\Lambda \text{CDM}}_\ell)\textbf{B}^T -\textbf{B}\text{Cov}(\vec{C}^{\Lambda \text{CDM}}_\ell, \vec{C}^o_\ell)\textbf{A}^T\\
    \label{eq:covariance_matrix}
    &-\textbf{A}\text{Cov}(\vec{C}^o_\ell, \vec{C}^{\text{fg}}_\ell)\textbf{Z}^T -\textbf{Z}\text{Cov}(\vec{C}^{\text{fg}}_\ell, \vec{C}^o_\ell )\textbf{A}^T,
\end{align}
where we have assumed that there are no correlations between $\Lambda\text{CDM}$ power spectra and the polarized foreground power spectra. Unfortunately, we do not have access to $C_\ell^{EB, \textrm{fg}}$ explicitly. In the next section, however, we show how we can use a filamentary dust model to relate the $EB$ spectra of the foreground to the $E$-modes and $B$-modes. In this work, we thus set $\vec{C}^{\text{fg}}_\ell=0$, and in the next section incorporate the effect of the foreground into the matrices $\textbf{A}$ and $\textbf{B}$. We are also not including the $\Lambda \text{CDM}$ power spectrum $\vec{C}^{\Lambda \text{CDM}}_\ell$ in the covariance matrix in Eq.~\eqref{eq:covariance_matrix} since we find a negligible difference in the posterior distribution when we include it. Thus, we only use $\textbf{M}_\ell = \textbf{A}\text{Cov}(\vec{C}^o_\ell, \vec{C}^o_\ell)\textbf{A}^T$.

As in previous works \citep{2022arXiv220107682D, Minami:2020odp}, we bin the spectra into groups of $\Delta \ell = 20$ multipoles to minimize noise and cosmic variance. We focus only on high-$\ell$ multipoles by using $\ell_\text{min} = 51$ and $\ell_\text{max} = 1490$, which gives the number of bins $N_{\text{bins}} = 72$. Explicitly, we get
\begin{align}
    C^{XY}_b &= \frac{1}{\Delta \ell} \sum_{\ell \in b} C^{XY}_{\ell}\\
     \text{Cov}(C^{XY}_b, C^{ZW}_b) &= \frac{1}{\Delta \ell ^ 2} \sum_{\ell \in b}  \text{Cov}(C^{XY}_\ell, C^{ZW}_\ell),
\end{align}
where $b$ is the bin number. To account for masks, we divide the covariance matrix, $\textbf{M}_b$, by the sky fraction $\fsky$ which we define as
\begin{equation}
    \fsky = \frac{1}{N_{\text{pix}}}\frac{\left(\sum^{N_{\text{pix}}}_{i=1}w_i^2\right)^2}{ \sum^{N_{\text{pix}}}_{i=1}w_i^4}.
\end{equation}
Here, $w_i$ is the weight of the apodized mask at pixel $i$ and $N_\text{pix}$ is the number of pixels \citep{Hivon:2002, Challinor:2004pr}.

We solve the equation by sampling over $\beta_i$ and $\alpha_i$ using a Markov chain Monte Carlo to evaluate:
\begin{equation}
    \ln L = -\frac{1}{2} \sum_{b=1}^{N_{\text{bin}}} \left(\vec{v}_b^T \textbf{M}_b^{-1}\vec{v}_b + \ln |\textbf{M}_b|\right),
\end{equation}
where $\textbf{M}_b$ is the binned covariance matrix. The latter term $\ln |\textbf{M}_b|$ in the log-likelihood was used in \cite{2022arXiv220107682D} but not in \cite{MinamiKomatsu:2020}. In simulations, we found that by adding the term, we obtained posteriors for the angles that were better aligned with the input values, especially for smaller values of $\fsky$.

To get the theoretical $\Lambda \text{CDM}$ power spectra of the CMB, we use CAMB\footnote{\url{https://github.com/cmbant/CAMB}} \citep{Lewis:2000} using the cosmological parameters from \planck\ 2018 \citep{planckcosmo:2018}. We then beam smooth the theoretical $\Lambda$CDM spectra so that
\begin{equation}
    \vec{C}^{\Lambda \text{CDM}}_\ell = \begin{bmatrix}
    C^{EE, \Lambda \text{CDM}}_{\ell}  W^{E_iE_j, E_iE_j}_\ell w^2_{\text{pix}, \ell}, \,C^{BB, \Lambda \text{CDM}}_{\ell} W^{B_iB_j, B_iB_j}_\ell w^2_{\text{pix}, \ell}
    \end{bmatrix},
\end{equation}
where $W^{XY, XY}_\ell$ are the beam window matrices and $w_{\text{pix}, \ell}$ are the pixel window functions. Both $i$ and $j$ denote the frequency and data split.

We first use five polarization-sensitive frequency bands from the \planck\ satellite. From the HFI, we adopt all the polarization-sensitive bands, namely the $100, 143, 217,$ and $353$\,GHz channels, along with the $70$\,GHz band from the LFI. The \npipe\, pipeline splits the data for each frequency band into two groups of maps, $A$ and $B$. For the 70\,GHz and higher frequency channels, these data splits are detector splits. This gives one average miscalibration angle for each set of detectors and, hence, we get ten miscalibration angles $\alpha_i$ where $i = 70\text{A}, ...,353\text{B}$. To get the observed power spectra $\vec{C}^o_{\ell}$ from these maps, we use \texttt{PolSpice}\footnote{\url{http://www2.iap.fr/users/hivon/software/PolSpice/}} \citep{Chon:2003gx}.

We use the same masks as in \cite{2022arXiv220107682D}. They all include a carbon monoxide (CO) mask that excludes pixels where there are bright CO emission lines. Unpolarized CO emission can cause intensity to polarization leakage, and so pixels with CO emission brighter than $45 \text{K}_{\text{RJ}}\,\text{km s}^{-1}$ are excluded. These masks also exclude point sources that are the union of the polarized HFI maps. Four Galactic masks that cover $5\%$, $10\%$, $20\%$, and $30\%$ are combined with the CO and point-source masks, leaving one mask with only a combined CO and point-source mask. This gives five masks with a sky coverage of $\fsky = 0.93, 0.90, 0.85, 0.75,$ and $0.63$.

We look at the frequency dependence of a possible cosmic birefringence angle $\beta_\nu$ where $\nu$ is the frequency of the photons. To test the frequency dependence of the birefringence angle, we will sample $\beta_\nu$ in two different ways. First, we sample $\beta_\nu$ for each frequency band individually for $\nu \in \{70, 100, 143, 217, 353\}$\,GHz. To minimize the loss of the signal-to-noise ratio, we also chose to fit the cosmic birefringence angle to a power law $\beta_\nu = \beta_0 (\nu / \nu_0)^n$. Here, $\nu_0$ is a reference frequency that we set to $150$\,GHz, since we find it to give the smallest error bars on $\beta_0$.

In the last section, we analyze the inclusion of the 30 and 44\,GHz frequency bands separately. This is due to their polarized foreground emission being dominated by synchrotron emission rather than dust emission. For the 30 and 44\,GHz bands, the data splits A and B are, in fact, time splits, unlike the other detectors which are detector splits. We, therefore, sample one miscalibration angle $\alpha_i$ for each of the two frequency bands. The other bands, 70\,GHz and higher frequency bands, still have two miscalibration angles each. Therefore, using all seven polarized \planck\ maps gives us a total of 12 miscalibration angles.

Even though the 70\,GHz maps contain more polarized synchrotron emission than dust emission, we made the choice of including the 70\,GHz channel in the analysis with the HFI bands. This is partly to see if the measured birefringence angle $\beta_\nu$ at 70\,GHz is also heavily mask-dependent, which is hypothesized to be due to $EB$ correlations caused by polarized dust emission which is sub-dominate at 70\,GHz. The 70\,GHz channel is also a band where there is a less total foreground contribution, and so, we might hope to get a stronger signal of $\beta_\nu$ at this frequency.

\section{Filament model for $EB$ correlations in dust}
\label{sec:filament}

The largest uncertainty of Eq. \eqref{eq:full_equation} is the intrinsic $EB$ correlations of the polarized foreground emission. Unlike the other power spectra, we do not have a reliable and robust theory of what this could be. However, as in \cite{2022arXiv220107682D} we make an ansatz based on the filamentary dust model of $EB$ from 
\cite{Clark:2021kze} as our best estimate of the intrinsic dust $EB$ correlations.

First, we start by relating the intrinsic effect of the foreground's $EB$ correlations to the miscalibration angle $\alpha$. By defining the correlation ratio as:
\begin{equation}
        r_\ell^{XY} \equiv \frac{C^{XY}_\ell}{\sqrt{C^{XX}_\ell C^{YY}_\ell}},
\end{equation}
we can write $C^{EB, \text{fg}}_\ell = r^{EB, \text{fg}}_\ell \sqrt{C^{EE, \text{fg}}_\ell C^{BB, \text{fg}}_\ell}$. The intrinsic $EE$ and $BB$-correlations of dust emission are approximately proportional to each other, $C^{BB, \text{fg}}_\ell = \xi C^{EE, \text{fg}}_\ell$ where $\xi \approx 0.5$ \citep{planckdust:2018}. This gives \citep{Minami:2019ruj}
\begin{equation}
    C^{EB, \text{fg}}_\ell = \frac{r^{EB, \text{fg}}_\ell\sqrt{\xi}}{1-\xi} \left(C^{EE, \text{fg}}_\ell - C^{BB, \text{fg}}_\ell\right).
\end{equation}
For the sake of argument, let us assume that the correlation ratio is independent of multipoles, $r^{EB, \text{fg}}_\ell = r^{EB, \text{fg}}$. Then we can define an effective angle $\gamma$, $\sin(4\gamma)/2 = r^{EB, \text{fg}} \sqrt{\xi}/(1-\xi)$. This angle is completely degenerate with the miscalibration angle $\alpha$. By ignoring the intrinsic $EB$ correlations of the foreground emission, instead of measuring $\alpha$ and $\beta$, we are measuring $\alpha + \gamma$ and $\beta - \gamma$, respectively, which leaves the sum of $\alpha + \beta$ as invariant.

Interstellar dust dominates the Galactic polarized emission at frequencies $\gtrapprox 100$\,GHz \citep{planckdust:2018}. The polarized dust emission was hypothesized in \cite{Huffenberger:2019mjx} to cause $EB$ correlations when the filaments have their long axis partially misaligned with interstellar magnetic fields. The misalignment angle can be characterized by a multipole-dependent effective angle $\psi^{\text{dust}}_\ell$ \citep{Clark:2021kze}. This angle works as an averaged misalignment angle for filaments on the given angular scale. The estimate for this angle was given as:
\begin{align}
    \psi^{\text{dust}}_\ell = \frac{1}{2}\arctan\left(\frac{C^{TB}_\ell}{C^{TE}_\ell  }\right).
\end{align}
With this angle, \cite{Clark:2021kze} estimated the $EB$ correlations to be
\begin{equation}
    C^{EB, \text{dust}}_\ell = |r_\ell^{TB, \text{dust}}| \sqrt{C^{EE, \text{dust}}_\ell C^{BB, \text{dust}}_\ell}\sin(4\psi^{\text{dust}}_\ell).
\end{equation}
Due to the approximate proportionality between $C^{EE, \text{dust}}_\ell$ and $C^{BB, \text{dust}}_\ell$, and to avoid taking the square root of noisy data, we set explicitly $C^{BB, \text{dust}}_\ell \rightarrow C^{EE, \text{dust}}_\ell$. We therefore use the ansatz from \cite{2022arXiv220107682D}:
\begin{equation}
     C^{EB, \text{dust}}_\ell = A_\ell C^{EE, \text{dust}}_\ell \sin \left(4\psi^{\text{dust}}_\ell\right),
\end{equation}
where $A_\ell$ is equivalent to $|r^{TB}_{\ell}|$ which we then sample over different multipole ranges along with $\alpha_i$ and $\beta_i$. For small angles, we can relate this to the effective angle $\gamma_\ell$:
\begin{equation}
    \label{eq:gamma_ell}
    \gamma_\ell \approx A_\ell \frac{C^{EE, \text{dust}}_\ell}{C^{EE, \text{dust}}_\ell - C^{BB, \text{dust}}_\ell} \frac{C^{TB, \text{dust}}_\ell}{C^{TE, \text{dust}}_\ell}.
\end{equation}

To get the spectra needed to compute $\gamma_\ell$, we take the cross-power spectrum of the $A$ and $B$ splits at $353$\,GHz with the respective mask. The polarization of the CMB emission is much less intense at this frequency, so it is expected that this frequency band describes polarized dust emission. Taking out $A_\ell$, we calculate Eq. \eqref{eq:gamma_ell} with the 353\,GHz spectra and apply it to all frequency bands. Since we are taking ratios of spectra, the spectral energy density of the power spectra cancels out for other frequency bands. Thus, we expect $\gamma_\ell$ to be independent of the frequency. This has been confirmed numerically by allowing $A_\ell$ to vary with frequency, which gives similar posterior distributions of the sampled parameters.

For the $70$\,GHz band, polarized thermal dust no longer dominates, and we expect synchrotron emission to be of similar or slightly higher amplitude than polarized dust emissions \citep{PlanckForeground:2014zja, PlanckComponentSeparation:2018, 2020arXiv201105609B}. Nevertheless, we still apply the filamentary dust model to the $70$\,GHz band because we still expect a dust contribution. As we comment on more later, synchrotron emission has not been found to contain any intrinsic $EB$ correlations \citep{Martire:2021gbc}, and so, we assume that polarized dust emission is the only foreground component that is expected to contain any intrinsic $EB$ for the 70\,GHz channel.

Similarly to \cite{2022arXiv220107682D}, when we use the filamentary dust model, we sample four multipole ranges for $A_\ell$ in combination with sampling over $\beta_\nu$ and $\alpha_i$. For $A_\ell$, we choose the $\ell$-ranges $51\leq\ell\leq130$, $131\leq \ell \leq 210$, $211\leq \ell \leq 510$, $511\leq \ell \leq 1490$.

We can then relate the intrinsic $EB$ of the foreground to the $EE$ and $BB$: $C^{E_iB_j, \textrm{fg}}_\ell = 2\gamma_\ell (C^{E_iE_j, \textrm{fg}}_\ell - C^{B_iB_j, \textrm{fg}}_\ell)$. In matrix notation, this becomes:
\begin{equation}
    \begin{bmatrix}
       C_\ell^{E_iB_j, \text{fg}}\\
           C_\ell^{B_iE_j, \text{fg}}
    \end{bmatrix} = \mathbf{F}_\ell \begin{bmatrix}
       C_\ell^{E_iE_j, \text{fg}}\\
          C_\ell^{B_iB_j, \text{fg}}
    \end{bmatrix}, \;\;\; \mathbf{F}_\ell = 2\gamma_\ell\begin{bmatrix}
        1 & -1\\
        1 & -1
    \end{bmatrix}.
\end{equation}
We can now put this expression into Eqs. \eqref{eq:c_ee_c_bb} and \eqref{eq:c_eb}, and we end up with modified $\textbf{A}$ and $\textbf{B}$ matrices \citep{2022arXiv220107682D}:
\begin{align}
    \mathbf{A}'_{\ell, ij} =& \left[- \Lambda_{\ell}^T(\alpha_i, \alpha_j)\mathbf{\Lambda}_{\ell}^{-1}(\alpha_i, \alpha_j),\, 1\right],
\end{align}
\begin{align}
    \nonumber
    \textbf{B}'_{\ell, ij}= \bigg[&R^T(\alpha_i+\beta_i,\alpha_j+\beta_j)-\Lambda_\ell^T(\alpha_i,\alpha_j)\mathbf{\Lambda}_\ell^{-1}(\alpha_i, \alpha_j)\\
    &\cdot\textbf{R}(\alpha_i+\beta_i, \alpha_j+\beta_j)\bigg],
\end{align}
where we have defined
\begin{align}
    \mathbf{\Lambda}_\ell(\alpha_i, \alpha_j) &= \mathbf{R}(\alpha_i, \alpha_j) + \mathbf{D}(\alpha_i, \alpha_j)\mathbf{F}_\ell,\\
    \Lambda_\ell^T(\alpha_i, \alpha_j) &= R^T(\alpha_i, \alpha_j) + D^T(\alpha_i, \alpha_j)\mathbf{F}_\ell.
\end{align}
\begin{figure}
\centering
\includegraphics[width=\linewidth]{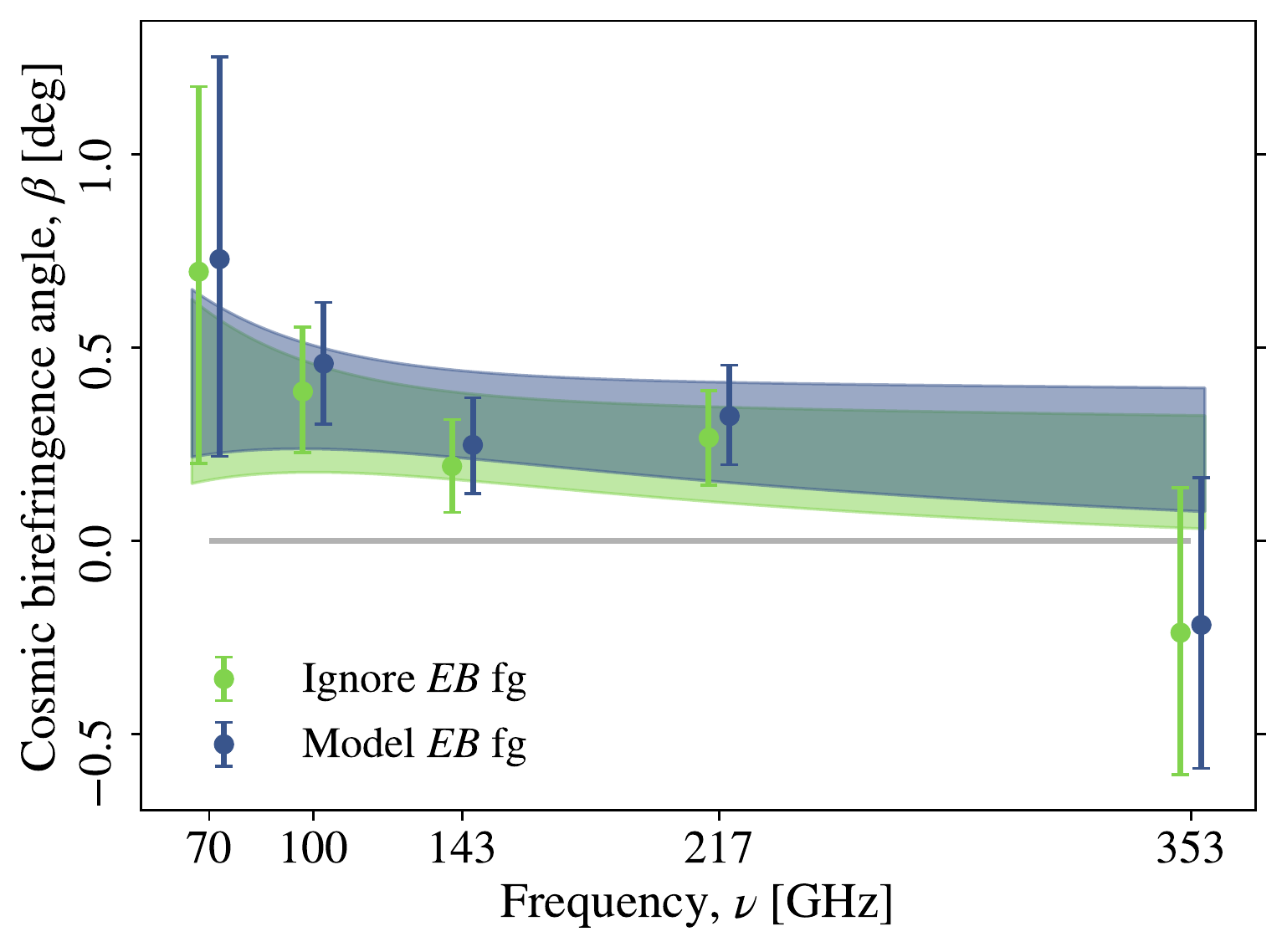}
\caption{Constraints on $\beta_\nu$ for each frequency band considered at nearly full-sky, $\fsky = 0.93$. The constraints of the power law measurement, $\beta_\nu=\beta_0(\nu/\nu_0)^n$, are shown as colored $1\sigma$-bands. The individual measurements of $\beta_\nu$ for each frequency channel are shown as error bars.}
\label{fig:nearly-full-sky}
\end{figure}
The equation we are solving for now is
\begin{equation}
    \textbf{A}'_{\ell}\vec{C}^o_\ell - \textbf{B}'_{\ell}\vec{C}^{\Lambda \text{CDM}}_\ell= 0,
\end{equation}
with the corresponding covariance matrix $\textbf{M}'_{\ell}$ using $\textbf{A}'_{\ell}$ and $\textbf{B}'_{\ell}$ in Eq. \eqref{eq:covariance_matrix} with $\mathbf{Z}=0$. As mentioned, we find no difference in the posterior distribution when including the \LCDM\, power spectra in the covariance, and therefore, we set $\textbf{M}'_\ell = \textbf{A}'\text{Cov}(\vec{C}^o_\ell, \vec{C}^o_\ell)\textbf{A}'^T$.

\section{Results}
\label{sec:results}

In this section, we report the results of the analysis on the HFI bands and the 70\,GHz band from the LFI. We first report the nearly full-sky, $\fsky = 0.93$, measurement of a frequency-independent $\beta$. When neglecting the intrinsic $EB$ of the polarized dust emission, we find $\beta=0.31^\circ \pm 0.10^\circ$. When using the filamentary dust model, we find $\beta=0.37^\circ \pm 0.11^\circ$. We note that these are slightly different from those reported in \cite{2022arXiv220107682D} since we are including the 70\,GHz channel.

\begin{table}
\centering
\begin{tabular}{c | c  |  c}
\hline\hline &  Ignore $EB$ fg & Model $EB$ fg\\
$\nu$ [GHz]& $\beta_\nu$  &$\beta_\nu$ \\ 
\hline
    \phantom{0}70 &  \phantom{$-$}$0.72^\circ \pm 0.50^\circ$&   \phantom{$-$}$0.76^\circ \pm 0.50^\circ$   \\
    100 & \phantom{$-$}$0.41^\circ \pm 0.16^\circ$&    \phantom{$-$}$0.47^\circ \pm 0.16^\circ$     \\
    143 &  \phantom{$-$}$0.20^\circ \pm 0.12^\circ$&   \phantom{$-$}$0.26^\circ \pm 0.12^\circ$  \\
    217 & \phantom{$-$}$0.28^\circ \pm 0.13^\circ$&   \phantom{$-$}$0.33^\circ\pm0.13^\circ$   \\
    353 &  $-0.25^\circ\pm0.38^\circ$&   
    $-0.22^\circ\pm0.38^\circ$   \\
\hline
\hline
\end{tabular}
\caption{Individual measurements of $\beta_\nu$ at nearly full-sky, $\fsky=0.93$, with and without modeling the intrinsic $EB$ of the polarized dust emission.}
\label{table:ind-measurement} 
\end{table}
\begin{figure}
\centering
\includegraphics[width=\linewidth]{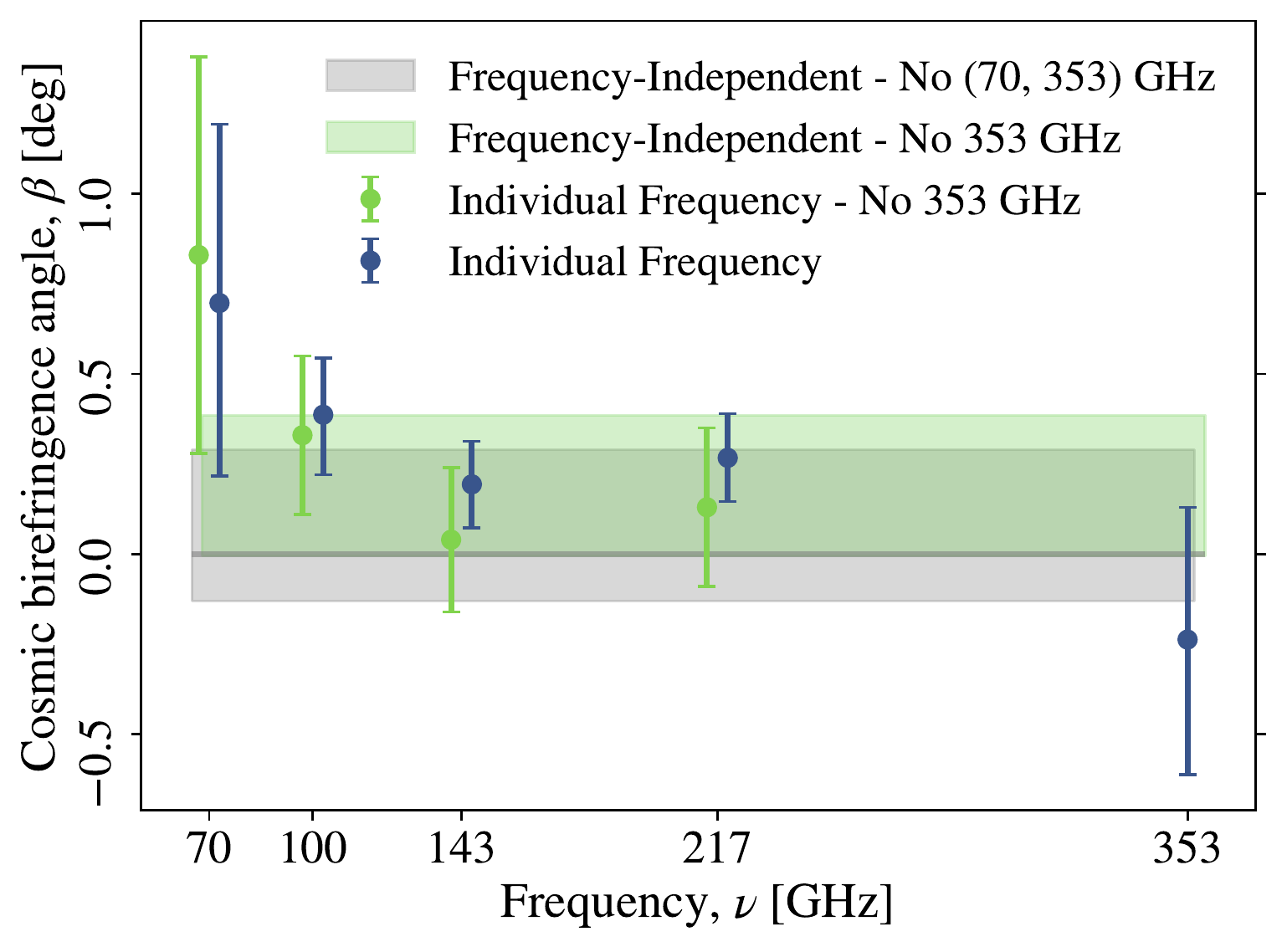}
\caption{Constraints on $\beta_\nu$ for each frequency band considered at nearly full-sky, $\fsky = 0.93$, when we ignore the $EB$ correlations of the foreground. The colored $1\sigma$-bands are frequency-independent measurements, unlike the power law model shown in Fig.\ref{fig:nearly-full-sky}.}
\label{fig:nearly-full-sky-353}
\end{figure}

For a frequency-dependent value of $\beta_\nu$, we show our nearly full-sky, $\fsky =0.93$, measurement in Fig. \ref{fig:nearly-full-sky}. This shows both the individual measurements of $\beta_\nu$ for each frequency band and the power law model, $\beta_\nu = \beta_0(\nu/\nu_0)^n$, where we use $\nu_0=150$\,GHz, since we find that this reference frequency gives the smallest uncertainty in $\beta_0$. The power law models are shown as $1\sigma$-bands, and we show the results both when using the filamentary model for the polarized dust emission and when we ignore the intrinsic $EB$ of the foreground emission.

For the individual measurements, we quote the nearly full-sky values in Table \ref{table:ind-measurement}, while for the power law model, we find $\beta_0=0.26^\circ \pm 0.11^\circ$ and $n=-0.45^{+0.61}_{-0.82}$ when ignoring $C_\ell^{EB, \textrm{fg}}$, and, otherwise, $\beta_0 = 0.33^\circ \pm 0.12^\circ$ and $n=-0.37^{+0.49}_{-0.64}$ when we use the filamentary dust model for $C_\ell^{EB, \textrm{fg}}$.

\begin{figure}
\centering
\begin{subfigure}{\linewidth}
\centering
\includegraphics[width=\linewidth]{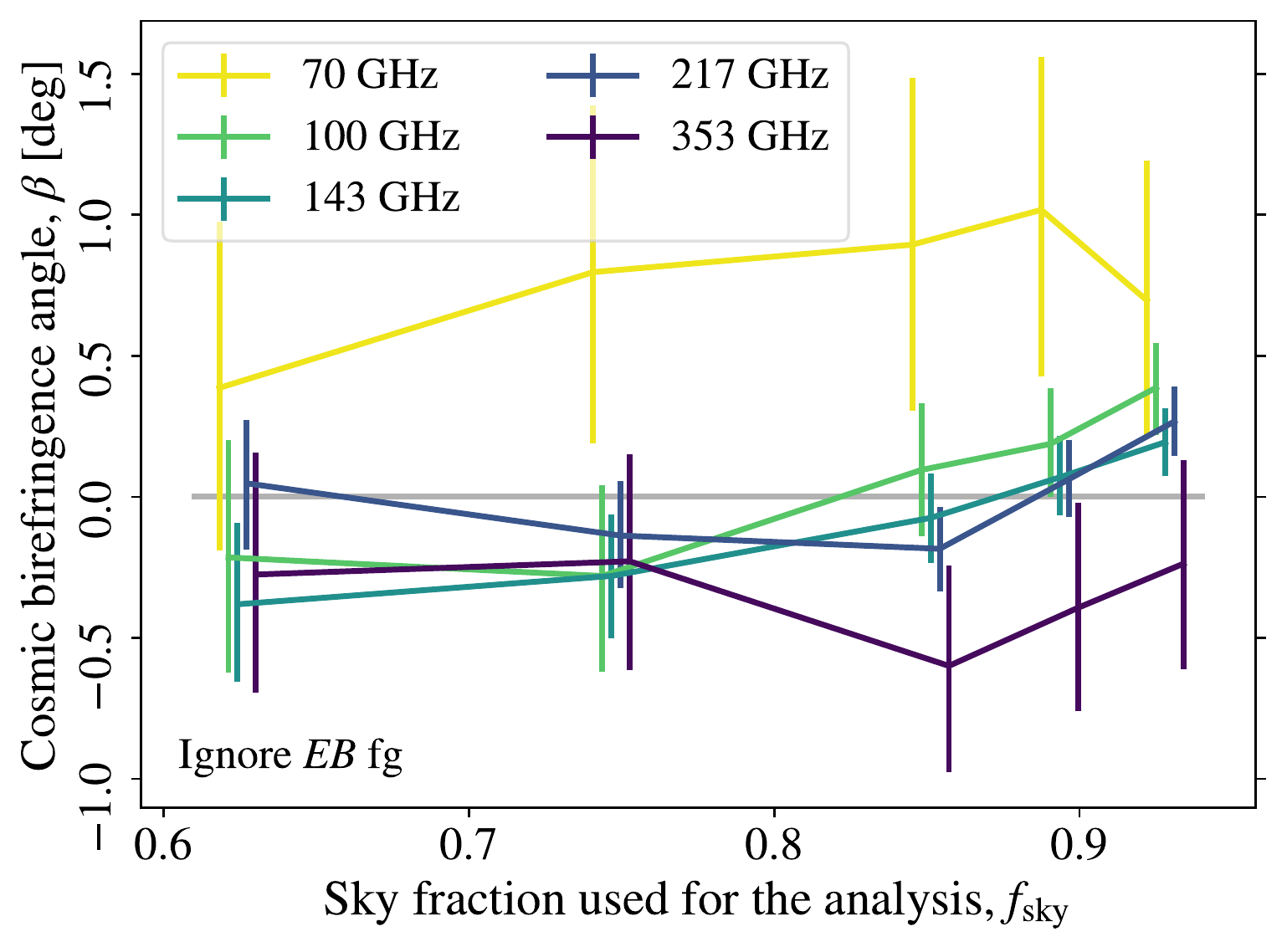}
\end{subfigure}

\begin{subfigure}{\linewidth}
\centering
\includegraphics[width=\linewidth]{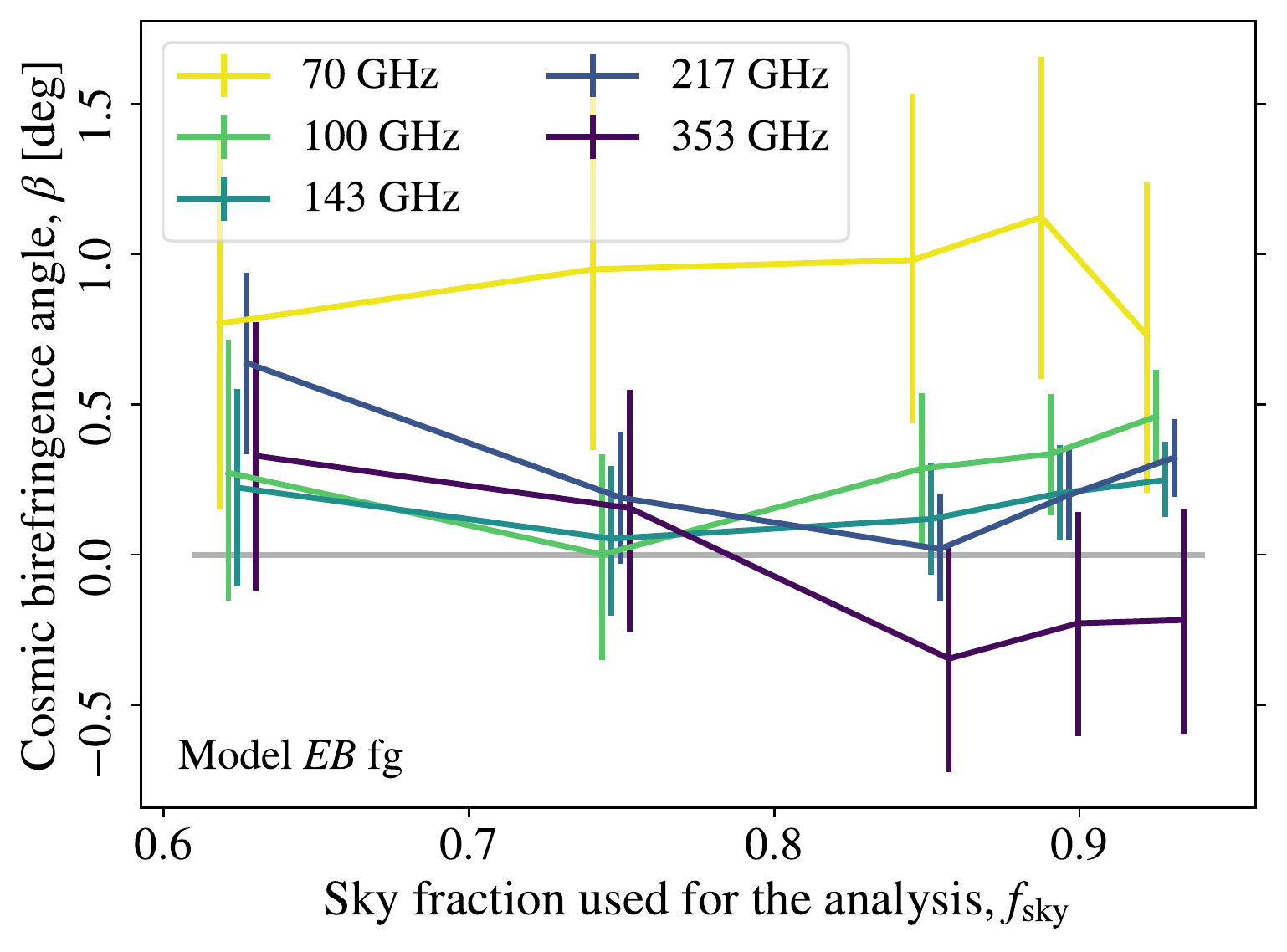}
\end{subfigure}
\caption{Constraints on a frequency-dependent $\beta$ for various values of $f_\mathrm{sky}$. The upper panel assumes no intrinsic $EB$ from the polarized foreground, while the lower panel shows the modeling of the foreground $EB$ using the filament model.}
\label{fig:beta_i-fsky}
\end{figure}

\begin{figure}
\centering
\begin{subfigure}{\linewidth}
\centering
\includegraphics[width=\linewidth]{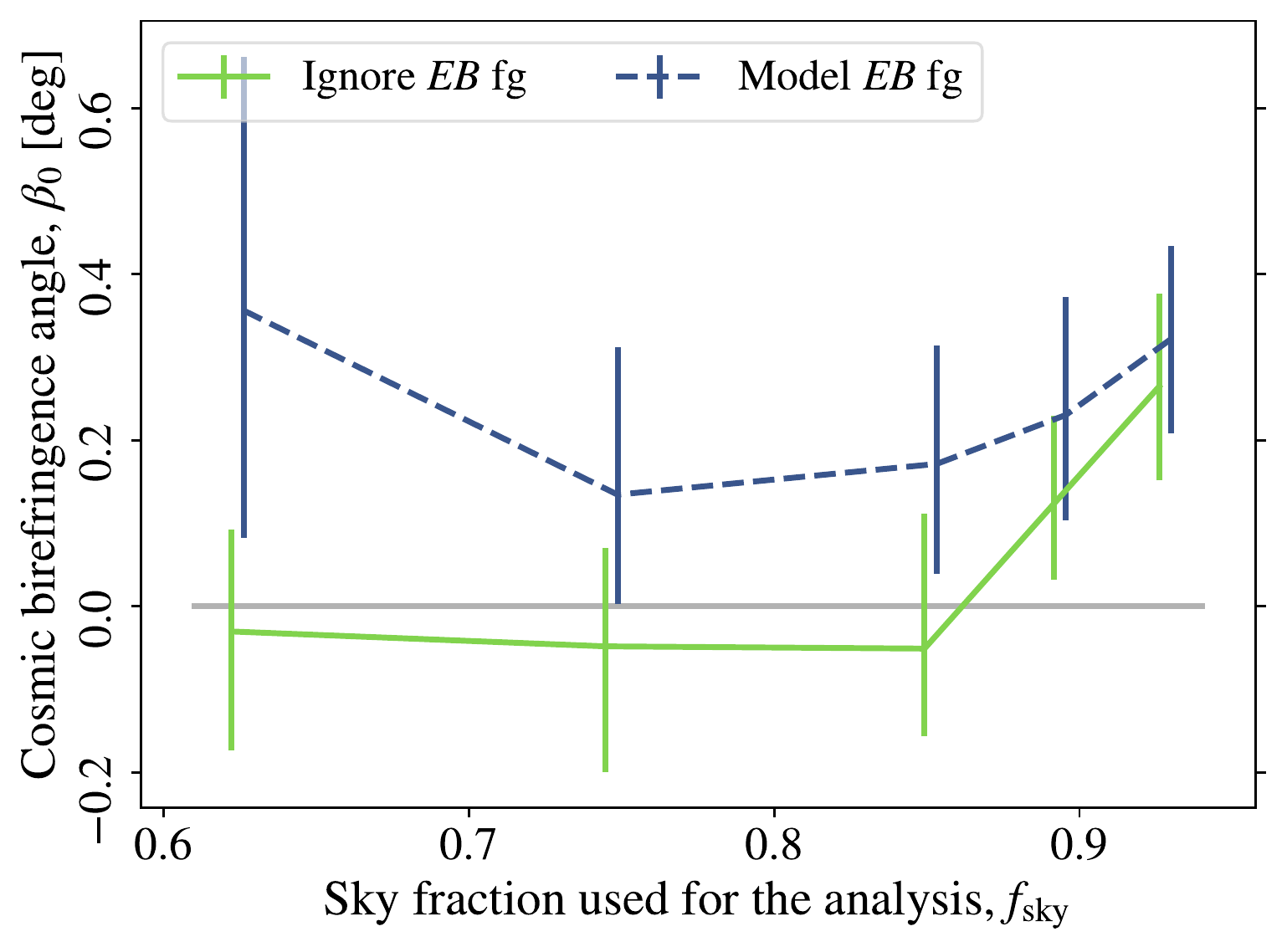}
\end{subfigure}

\begin{subfigure}{\linewidth}
\centering
\includegraphics[width=\linewidth]{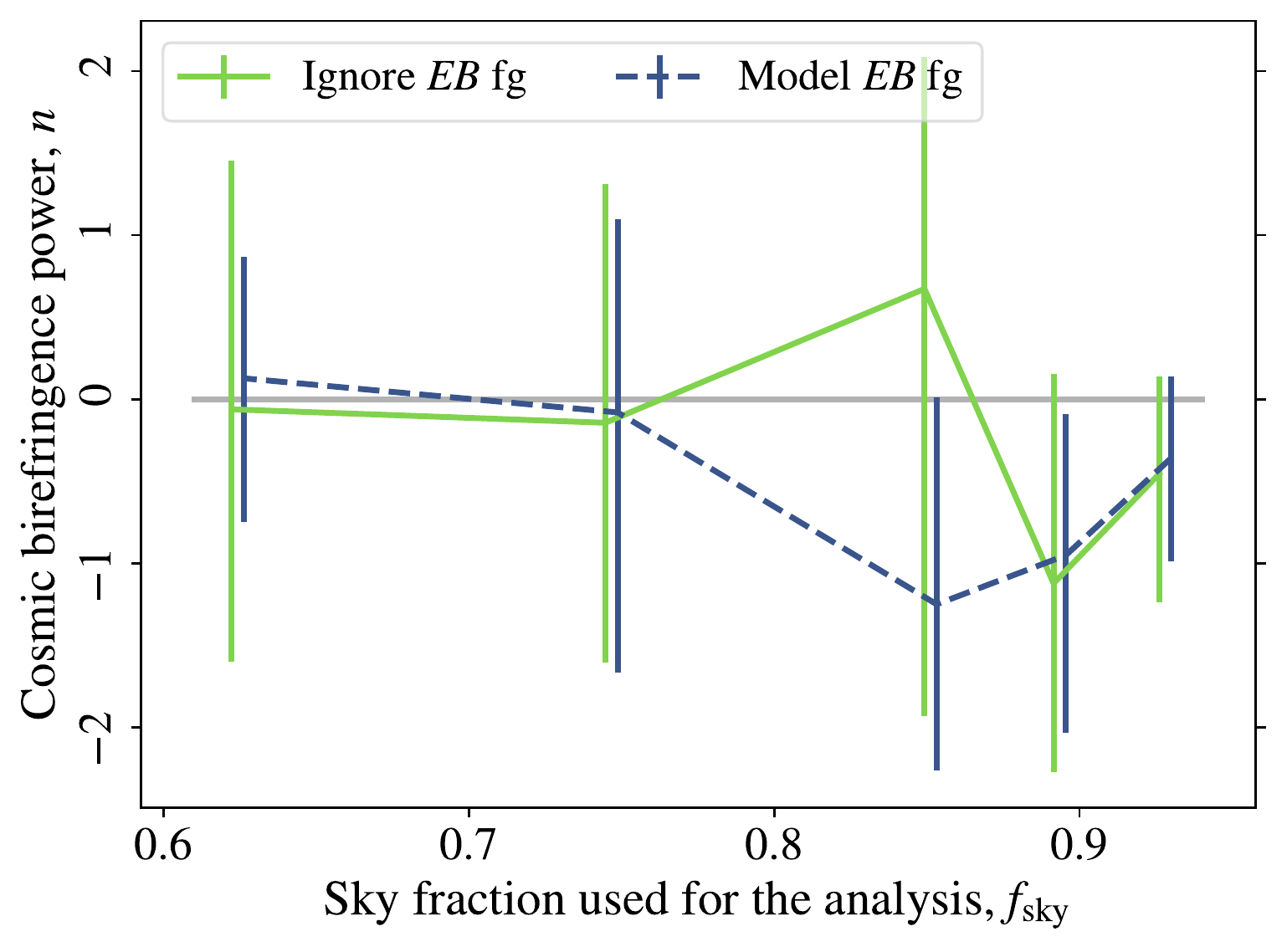}
\end{subfigure}
\caption{Measurement of the power law, $\beta_\nu = \beta_0(\nu/\nu_0)^n$, for various values of $f_\mathrm{sky}$. The upper panel shows the measurement of $\beta_0$, while the lower panel models the the observed $n$. We set a flat prior, $|n|\leq 3$.}
\label{fig:beta_0_n-fsky}
\end{figure}

It was reported in \cite{2022arXiv220107682D} that removing the 353\,GHz maps from the analysis gave a drop in the measurement of the frequency-independent birefringence angle, which suggests that the 353\,GHz band increases the measured value of $\beta$. On the other hand, we note that the individual measurement of $\beta_\nu$ for the 353\,GHz channel yields a negative value as seen in Fig.~\ref{fig:nearly-full-sky}. This leaves the role of the 353\,GHz channel unclear. Therefore, we investigate the effect of the 353\,GHz frequency band on the $\beta_\nu$ measurements in Fig.~\ref{fig:nearly-full-sky-353} where we ignore the intrinsic $EB$ power spectra of the foreground emission. The $1\sigma$-bands are in this case the frequency-independent measurements, where the gray band is the constraint taken from \cite{2022arXiv220107682D} when excluding both the 70 and 353\,GHz channels ($\beta = 0.08^\circ\pm0.21^\circ$) and the light green $1\sigma$-band excludes only the 353\,GHz channel ($\beta = 0.19^\circ \pm 0.19^\circ$). We also show the measurement of $\beta_\nu$ when we sample it individually for each frequency band. The figure indicates that the 353\,GHz band does not bias the measurement of the other bands significantly. Instead, it decreases the individual measurement uncertainties for all frequency bands. This can be explained by the domination of polarized dust emission at 353\,GHz which helps break the degeneracy between $\alpha$ and $\beta$ through the cross-power spectra with lower frequency channels.

The measured angles of a frequency-dependent $\beta_{\nu}$ for ${\nu \in \{70, 100, 143, 217, 353\}}$\,GHz are shown in Fig. \ref{fig:beta_i-fsky} as a function of the fractional sky coverage $\fsky$. The upper panel shows the probability distribution for the cosmic birefringence angle $\beta_\nu$ for each frequency band when we ignore the intrinsic $EB$ correlations of the foreground. We observe a decreasing $\beta_{\nu}$ as we mask more of the Galactic plane for $100$, $143$, and $217$\,GHz. This is consistent with \cite{2022arXiv220107682D} where a similar drop was shown for a single frequency-independent $\beta$. We do not see this decline for the 70\,GHz band. This band has a higher contribution from polarized synchrotron emission than the other bands, therefore, the mask-dependent contribution to $C_\ell^{EB, \textrm{dust}}$ is less significant. In the lower panel of Fig. \ref{fig:beta_i-fsky}, we show the probability distribution where we instead use the filament model for $C_\ell^{EB, \textrm{dust}}$. The drop in $\beta_\nu$ when we ignore the foreground $EB$ is mitigated when using this model. This reinforces the hypothesis that the intrinsic $EB$-correlation of dust is mostly positive for large Galactic masks as reported by \cite{Clark:2021kze}.

For the power law measurement, $\beta_\nu=\beta_0(\nu/\nu_0)^n$, we show our result in Fig. \ref{fig:beta_0_n-fsky} as a function of the sky fraction, $\fsky$. When $\beta_0=0$, $n$ is fully degenerate. Since we find no theoretical models for large or small spectral indices, $n$, and no support from the individual measurements of $\beta_\nu$ that there should be a large spectral index $n$ dependence, we set a flat prior $|n| \leq 3$. This prevents $n$ from diverging around $\beta_0\approx 0$. When we remove the prior on $n$, we find no support for other solutions at higher $|n|$.

The decline in the cosmic birefringence angle parameter $\beta_0$ is also apparent at large Galactic masks when the intrinsic $EB$ correlations of the polarized foreground emission are ignored. This drop is mostly mitigated when the filamentary dust model for $EB$ is taken into account.

\begin{table}
\centering
\begin{tabular}{c | c c | c c}
\hline\hline &  \multicolumn{2}{c|}{Ignore $EB$ fg} &   \multicolumn{2}{c}{Model $EB$ fg}\\
$n$& $\beta_0$ & $\Delta \chi^2$ &$\beta_0$& $\Delta \chi^2$\\ 
\hline
    \phantom{$-$}$2$ &  $0.03^\circ \pm 0.05^\circ$   & 8.21 & $0.04^\circ \pm 0.05^\circ$  &9.45  \\
    \phantom{$-$}$1$ & $0.17^\circ \pm 0.08^\circ$    & 4.67  &  $0.21^\circ \pm 0.09^\circ$   & 5.60  \\
    \phantom{$-$}$0$ &  $0.31^\circ \pm 0.10^\circ$     & 0.00   & $0.37^\circ \pm 0.11^\circ$  & 0.00\\
    $-1$ & $0.25^\circ \pm 0.09^\circ$     & 0.38  & $0.28^\circ\pm0.09^\circ$  & 0.75\\
    $-2$ &  $0.14^\circ\pm0.06^\circ$     & 2.25  & $0.15^\circ\pm0.06^\circ$  & 3.01\\
\hline
\hline
\end{tabular}
\caption{Measurement of $\beta_0$ for fixed values of $n$ for the power law measurement of the cosmic birefringence angle, $\beta_\nu=\beta_0(\nu/\nu_0)^n$ for $\nu_0=150$\,GHz. The corresponding $\Delta \chi^2$ is given with respect to $n=0$.}
\label{table:fixed-n} 
\end{table}
\begin{figure}
\centering
\includegraphics[width=\linewidth]{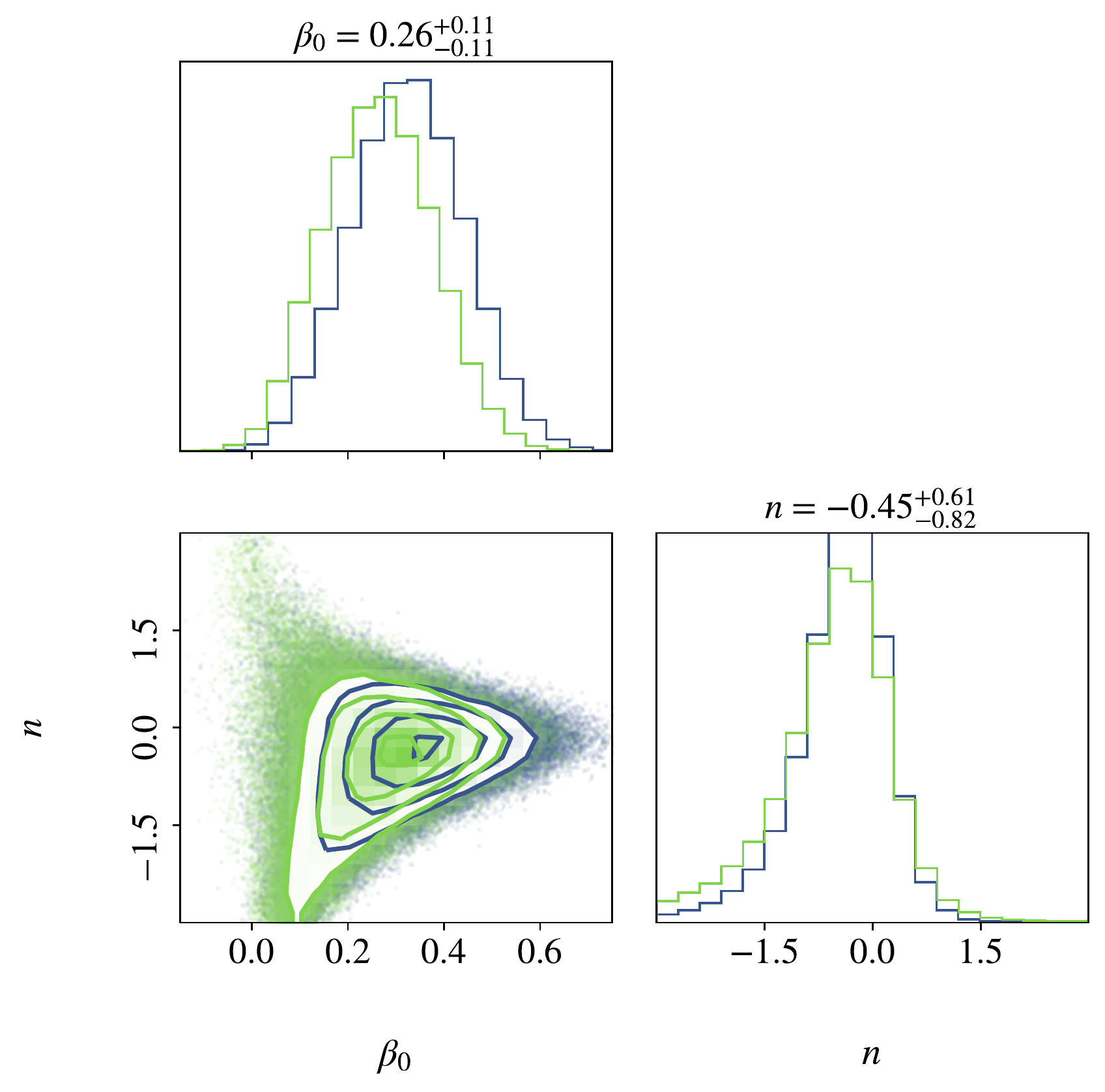}
\caption{Constraints on the power law model $\beta_\nu = \beta_0(\nu/\nu_0)^n$ for nearly full-sky $\fsky = 0.93$. The green contour shows the probability distribution when we ignore the intrinsic $EB$ of the foreground, while the blue contour shows that with the filament model for the foreground. The quoted values are for ignoring the $EB$ correlations of the foreground. The equivalent values using the filament model for dust $EB$ are $\beta_0 = 0.33^\circ \pm 0.12^\circ$ and $n=-0.37^{+0.49}_{-0.64}$.}
\label{fig:corner}
\end{figure}

The spectral index $n$ is shown in the lower panel of Fig.~\ref{fig:beta_0_n-fsky}. When we ignore $C^{EB, \textrm{fg}}_\ell$, we get results that are consistent with $\beta_0=0^\circ$ for larger sky cuts, and this generates broad uncertainties in $n$. The filamentary dust model yields a positive $\beta_0$, which allows us to probe the frequency dependence of the signal more accurately. For both nearly full-sky and large-sky cuts, we get a measurement that is consistent with a frequency-independent birefringence angle. The 68\%~C.L. lower limits for $\fsky = 0.9$ and $\fsky = 0.85$, touch $n=-2$, which would be expected from Faraday rotation caused by local magnetic fields.

The theoretical models considered in this work are based on integer values of $n$. Therefore, for nearly full-sky, $\fsky = 0.93$, we fix $n$ to be the integer values,  $n \in \{-2, -1, 0, 1, 2\}$, and only measure $\beta_0$ in Table \ref{table:fixed-n}. We also give the corresponding ${\Delta\chi^2 \equiv -2\ln L}$ with respect to $n=0$, which shows that the frequency-invariant solution $n=0$ is the favored integer solution.

In Fig. \ref{fig:corner}, we have included the posterior distribution of our power law assumption of $\beta$ in the $(n,\beta_0)$-plane at nearly full-sky, $\fsky = 0.93$. The blue contour shows the probability distribution when we use the filamentary dust model for the $EB$ foreground, while the green contour shows our measurement when we ignore the dust's intrinsic $EB$. We see that when $\beta_0$ is close to zero, $n$ is degenerate and seems to give $n<0$. However, for larger values of $\beta_0$, we see support for a frequency-independent birefringence angle, $n=0$.

This work is focused on probing the frequency dependence of $\beta_\nu$ assuming that the signal exists. Therefore, we check what constraints we get on $n$ by having a flat prior of $\beta_0 \geq 0.1^\circ$. In that case, we find $n=-0.43^{+0.58}_{-0.73}$ when ignoring $EB$ of the foreground emission and $n=-0.34^{+0.48}_{-0.56}$ when using the filament model.

\section{Including all the \planck\ polarized frequency maps}
\label{sec:all_maps}

\begin{figure}
\centering
\includegraphics[width=\linewidth]{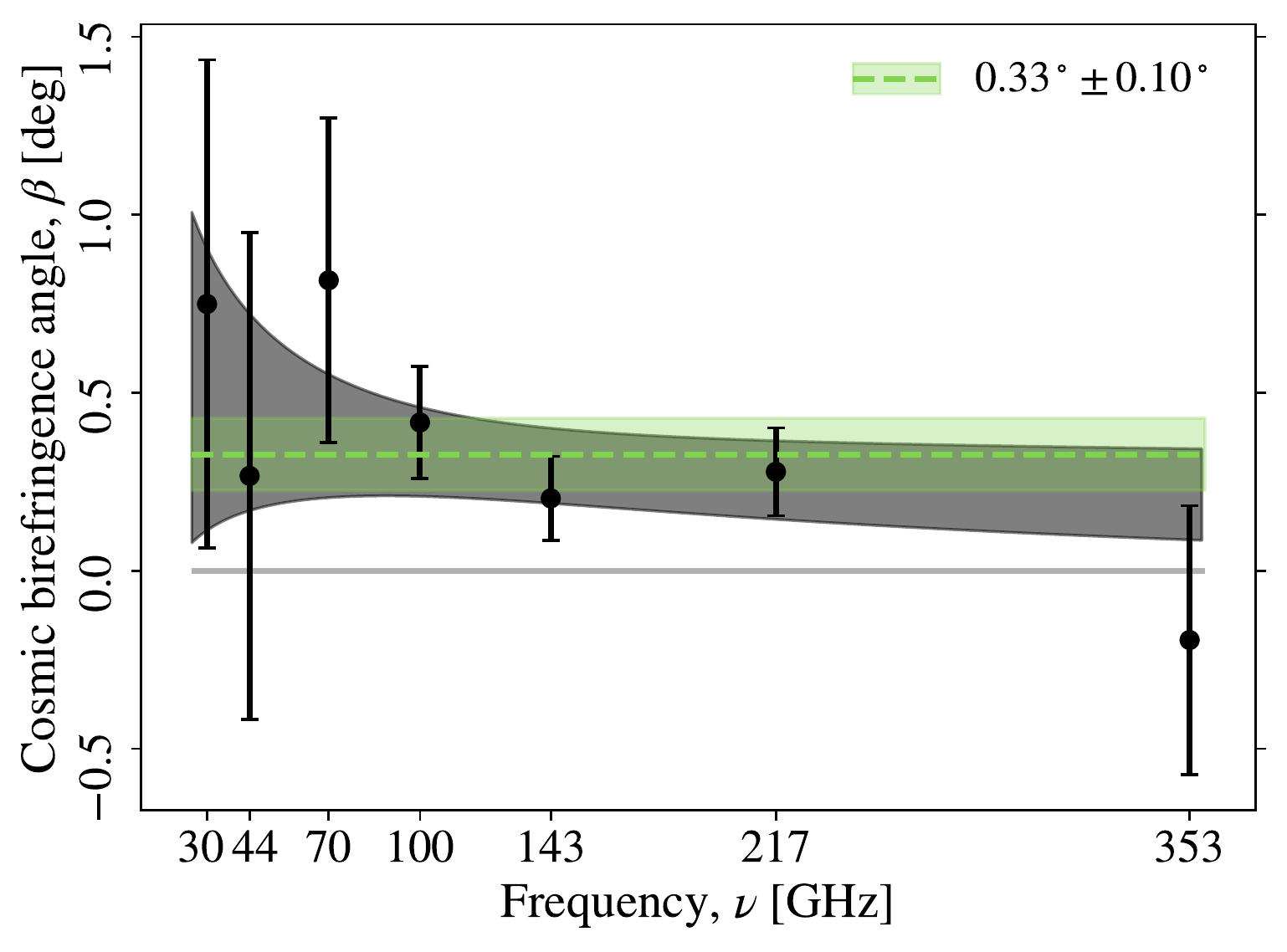}
\caption{Constraints on $\beta_\nu$ at nearly full-sky $\fsky = 0.92$, where we have included the 30 and 44\,GHz maps from the LFI. We ignore the intrinsic $EB$ of the foreground emission, and the $1\sigma$-band shows the constraints from the power law measurement, $\beta_\nu = \beta_0(\nu/\nu_0)^n$, where $\nu_0=150$\,GHz. The dashed green line with the $1\sigma$-band shows the frequency-independent $\beta$.}
\label{fig:fullsky_30_353}
\end{figure}

In this section, we include the 30 and 44\,GHz frequency channels from the LFI into our analysis. Thus, we are using all the polarized frequency maps from both the LFI and HFI of \planck. We treat the inclusion of these two frequency bands separately since they have a negligible contribution of polarized dust emission compared to the polarized synchrotron emission.

The $EB$ power spectrum of synchrotron emission has not been detected yet. \citet{Martire:2021gbc} found that the $EB$ of synchrotron emission is compatible with zero at $1\sigma$ using the 30GHz band of \planck\ and the \WMAP\ K-band for different sky fractions. The low signal-to-noise of these instruments puts weak constraints on it, but QUIJOTE and C-BASS aim to get a better understanding of the polarized synchrotron emission \citep{QUIJOTE:2018ntj, C-BASS:2018dwc}. Hence, we treat the full analysis of all \planck\ polarized maps with the inclusion of the 30 and 44\,GHz channels separately, and we do not include any modeling of the intrinsic $EB$ of either dust or synchrotron emission.

When including the two additional bands, we create a new point-source mask that is the union of all the LFI and HFI point-source masks. We combine it with the same CO and Galactic masks, and the new sky coverages become ${\fsky = 0.92, 0.88, 0.84, 0.74,}$ and $0.62$.

\begin{table}
\centering
\begin{tabular}{c | c}
\hline\hline &  Ignore $EB$ fg\\
$\nu$ [GHz]& $\beta_\nu$  \\ 
\hline
 \phantom{0}30 &  \phantom{$-$}$0.75^\circ \pm 0.68^\circ$ \\
  \phantom{0}44 &  \phantom{$-$}$0.27^\circ \pm 0.68^\circ$ \\
    \phantom{0}70 &  \phantom{$-$}$0.82^\circ \pm 0.45^\circ$ \\
    100 & \phantom{$-$}$0.42^\circ \pm 0.16^\circ$\\
    143 &  \phantom{$-$}$0.20^\circ \pm 0.12^\circ$\\
    217 &\phantom{$-$}$0.28^\circ \pm 0.12^\circ$\\
    353 &  $-0.19^\circ\pm0.38^\circ$\\
\hline
\hline
\end{tabular}
\caption{Individual measurements of $\beta_\nu$ at nearly full-sky, $\fsky=0.92$, without modeling the intrinsic $EB$ of the polarized dust emission.}
\label{table:ind-measurement_30_353} 
\end{table}

\begin{figure}
\centering
\includegraphics[width=\linewidth]{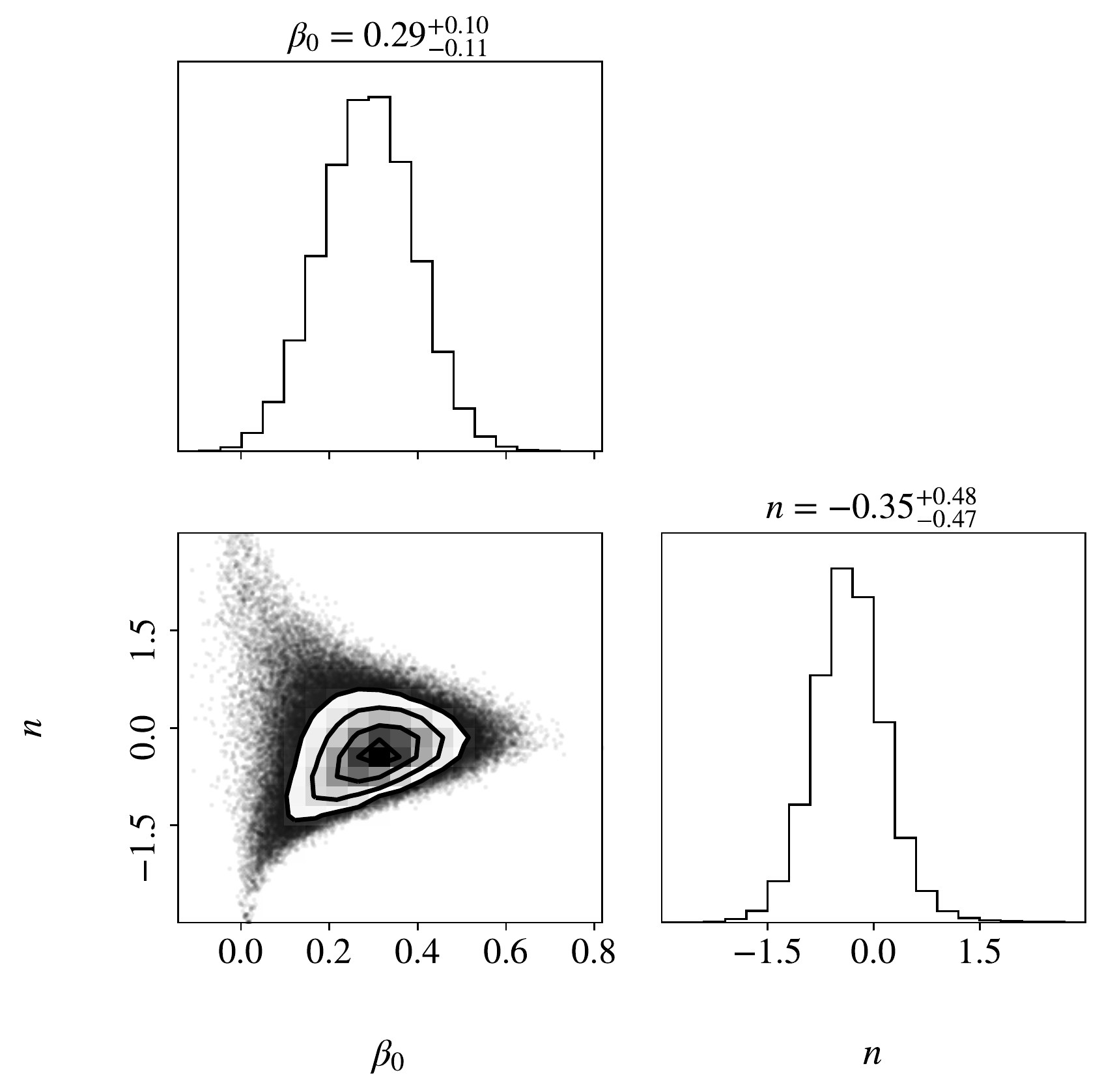}
\caption{Posterior distributions of $n$ and $\beta_0$ at nearly full-sky ${\fsky = 0.92}$ for the power law measurement, $\beta_\nu=\beta_0(\nu/\nu_0)^n$, where $\nu_0=150$\,GHz. We included the 30 and 44\,GHz maps from the LFI, and we ignore the intrinsic $EB$ of the foreground emission.}
\label{fig:corner_30_353}
\end{figure}

We present the nearly full-sky, $\fsky=0.92$, measurements in Fig.~\ref{fig:fullsky_30_353}. Here, we show the individual samples of $\beta_\nu$ for each frequency band, in addition to the power law measurement, $\beta_\nu=\beta_0(\nu/\nu_0)^n$, with $\nu_0=150$\,GHz shown as a $1\sigma$-band. The values of the individual measurement are quoted in Table \ref{table:ind-measurement_30_353}. For the power law model, we find $\beta_0=0.29^{\circ+0.10^\circ}_{\phantom{\circ}-0.11^\circ}$ and $n=-0.35^{+0.48}_{-0.47}$. The posterior distribution for this measurement is shown in Fig.~\ref{fig:corner_30_353}. We see that the individual measurements of $\beta_\nu$ at $30$ and $44$\,GHz bands are positive, and they slightly increase the measured value of $\beta_0$ for the power law measurement as seen in Fig.~\ref{fig:corner_30_353} compared to Fig.~\ref{fig:corner}. The inclusion of the two lowest frequency polarized LFI bands also help constrain better the spectral index $n$. The posterior distribution is slightly more centered around $n=0$ with smaller uncertainties.

\begin{figure}
\centering
\includegraphics[width=\linewidth]{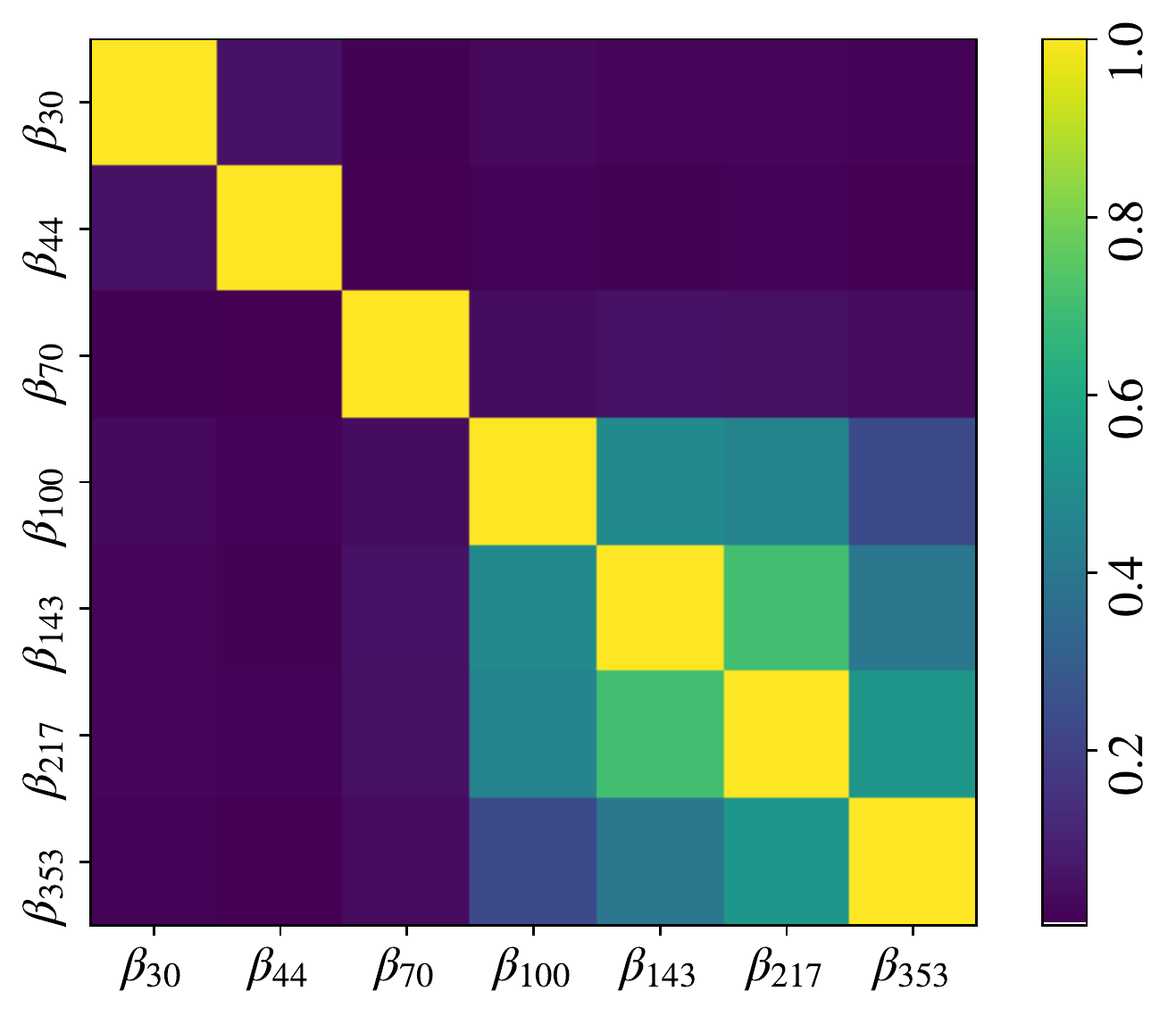}
\caption{Correlation coefficients between the individually sampled birefringence angles $\beta_\nu$ at nearly full-sky, ${\fsky=0.92}$.}
\label{fig:corr_beta_nu}
\end{figure}

The correlation coefficient matrix for the sampled angles $\beta_\nu$ at nearly full-sky is shown in Fig.~\ref{fig:corr_beta_nu}. We see little correlations between each LFI band and other frequencies, but the HFI channels give correlated measurements of $\beta_\nu$. The strong correlations between $\beta_{\nu}$ for $\nu\in\{100, 143, 217\}$\,GHz could be due to the high signal-to-noise of $C^{\textrm{CMB}, EE}_\ell$ for these bands. The CMB $EE$ power spectrum is the same for all channels, and it strongly constrains the degeneracy $\alpha+\beta$. Through cross-correlations with the 353\,GHz band, we also get correlations with $\beta_{353}$. A combination of the low signal-to-noise ratio of $C_\ell^{EE}$ and the foreground being synchrotron rather than dust at LFI could explain the lack of correlation between the LFI and HFI channels. To get a better understanding of the correlation coefficients, we could analyze simulations to see if the lower signal-to-noise level in the LFI bands are the cause of the smaller correlation coefficients as compared to the HFI bands. However, we leave such an analysis aside for future work.

We also try to fit the parameters of the power law model by using the covariance matrix of $\beta_\nu$ and the average measurements found in Table~\ref{table:ind-measurement_30_353}. We retrieve the values $\beta_0=0.29^\circ \pm 0.11^\circ$ and $n=-0.36^{+0.48}_{-0.47}$, and we find that the shape of the posterior distribution is very consistent with the full $(\beta_0, n)$ analysis seen in Fig.~\ref{fig:corner_30_353}.

\begin{figure}
\centering
\includegraphics[width=\linewidth]{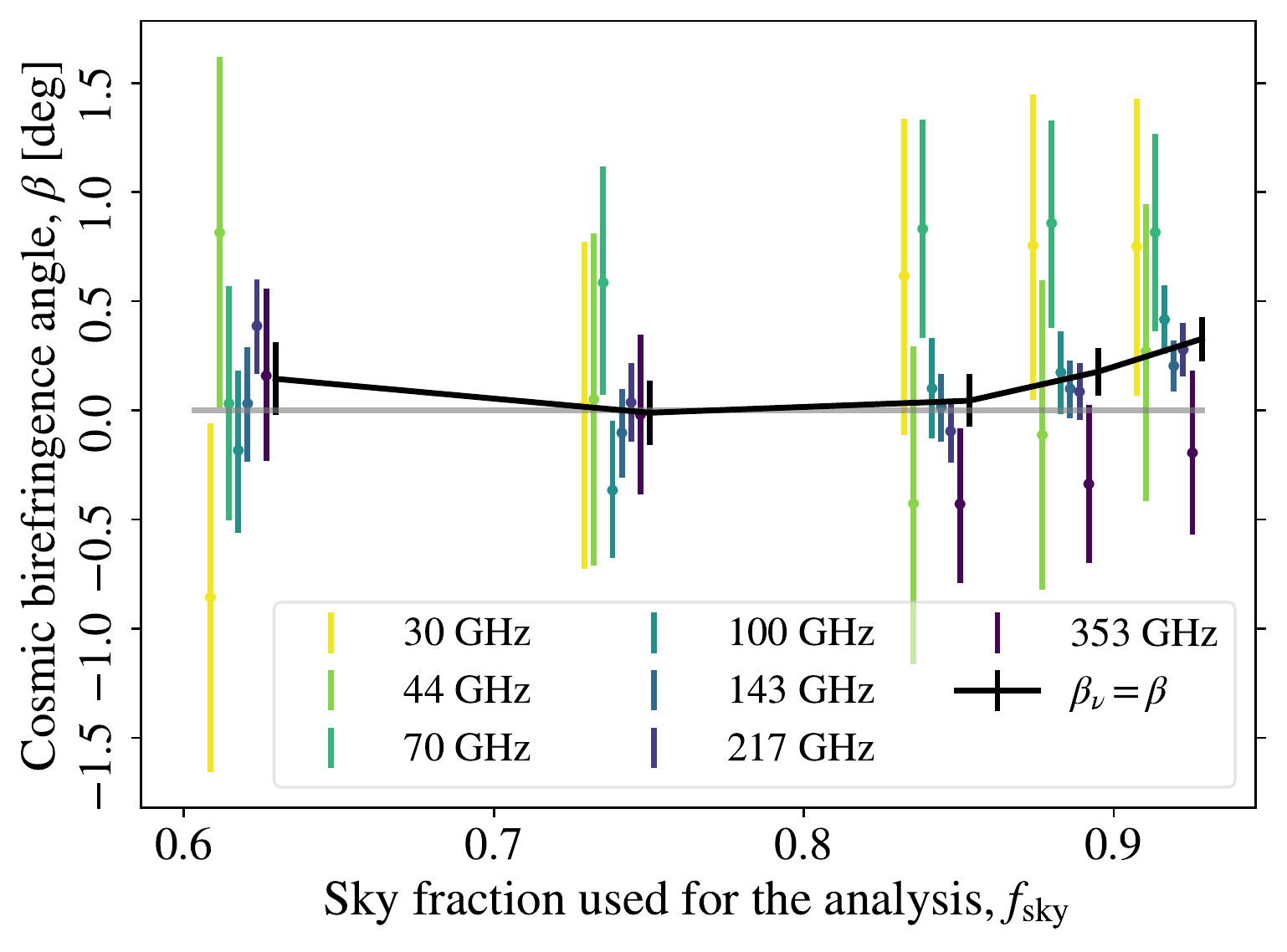}
\caption{Constraints on $\beta_\nu$ when we sample them individually for each frequency band. We have included the 30 and 44\,GHz maps from the LFI, and we ignore the intrinsic $EB$ of the foreground emission. The black line shows $\beta$ when we instead assume it is independent of frequency.}
\label{fig:ind_30_353}
\end{figure}

We show $\beta_\nu$ as a function of the sky coverage, $\fsky$, in Fig.~\ref{fig:ind_30_353}. This plot also includes the measurement when we assume no frequency dependence of $\beta$ shown as the black line. At nearly full-sky, we measure this frequency-independent angle to be $\beta=0.33^\circ \pm 0.10^\circ$. This figure can be compared to Fig.~1 in \citet{2022arXiv220107682D}. The drop in the frequency-independent $\beta$ as we mask more of the Galactic plane is partially mitigated by the inclusion of the LFI frequency bands. We also note that the individual measurements of $\beta_\nu$ for the HFI bands are generally lifted by the inclusion of $30$ and $44$\,GHz bands as can be seen when comparing with Fig.~\ref{fig:beta_i-fsky}. This is explained by the inclusion of a large number of new combinations of cross-power spectra with the $30$ and $44$\,GHz channels. This supports the hypothesis that the $EB$ power spectra of dust emission create a positive effective angle $\gamma$ at large Galactic masks that gives a lower measured value of $\beta$. The $30$ and $44$\,GHz should have a negligible contribution of dust compared to synchrotron emission.

\section{Conclusions}
\label{sec:conclusions}
In this work, we employ one of the pipelines of \cite{2022arXiv220107682D} to explore the possibility of a frequency-dependent cosmic birefringence angle $\beta_\nu$. First, we analyzed the HFI polarized frequency bands with the 70\,GHz band from the LFI. At nearly full-sky we get a positive measurement of $\beta_\nu$ for all individual frequency bands except $353$\,GHz. As expected from earlier investigations \citep{2022arXiv220107682D}, the measured cosmic birefringence angle declines as a function of larger Galactic masks when we set the intrinsic $EB$ correlations of the polarized foreground emission to zero. By using the $EB$ dust ansatz of \cite{2022arXiv220107682D}, motivated by the filamentary dust model of \cite{Clark:2021kze}, we can mitigate this effect.

We also sample a power law function of the birefringence angle, $\beta_\nu=\beta_0(\nu/\nu_0)^n$. Using both the filamentary dust model and neglecting the intrinsic dust $EB$, we get measurements that are inconsistent with a positive integer solution to the spectral index $n$, especially at large sky fractions. Some measurements at large sky coverages allow negative integer $n$ solutions. However, we get the smallest uncertainty on $n$ at $\fsky=0.93$, which favors the frequency-independent solution, $n=0$, over other integer solutions. 

We then included the 30 and 44\,GHz channels, thereby running the analysis on all the polarized \planck\ maps from both the LFI and HFI. These lowest frequency channels are dominated by polarized synchrotron emission. Measurements of the polarization of synchrotron emission indicate that the $EB$ correlations must be small and is so far consistent with zero \citep{Martire:2021gbc}. So we did not try to model the intrinsic $EB$ of the foreground emission.

The inclusion of all the LFI bands increased the mean value and tightened the constraints on the frequency-independent measurement of $\beta$ at all sky fractions considered as compared to \cite{2022arXiv220107682D} which only used the HFI channels. Especially at large Galactic cuts, we find that including the 30 and 44\,GHz bands raise the individual measurements of $\beta_\nu$ for the HFI bands. This supports the hypothesis of \citet{2022arXiv220107682D} and \citet{Clark:2021kze} that large Galactic masks give a generally positive intrinsic $EB$ of the polarized dust emission that biases the measurement of $\beta$ to a lower value. The contribution of polarized dust emission at 30 and 44\,GHz are negligible compared to synchrotron emission.

With all the polarized \planck\ maps, we measured ${\beta = 0.33^\circ \pm 0.10^\circ}$ assuming $\beta$ has no frequency dependence. We also sampled the parameters of the power law model, ${\beta_\nu=\beta_0(\nu/\nu_0)^n}$, at nearly full-sky for all the polarized LFI and HFI bands, and we found that the spectral index $n$ is more constrained by adding the 30 and 44\,GHz channels. The data favors $n=0$ as the integer solution and does not favor Faraday rotation as the cause of the potential cosmic birefringence signal.

\cite{2022arXiv220107682D} did a thorough investigation of the instrumental systematics of the HFI frequency bands by looking at realistic simulations of the \npipe\ processing. The impact of these systematics was found to be negligible for the measurement of $\beta$. However, such an analysis was not done for the LFI channels. This work has not analyzed the instrumental systematics of the LFI bands, and we leave this for future studies.

If the intrinsic $EB$ of polarized dust emission was the cause of our positive measurement of $\beta_\nu$, we would also have found a frequency-independent signal, as explained in \citet{2022arXiv220107682D}. However, dust emission is not the dominant foreground contribution for the LFI channels; synchrotron is. Dust $EB$ would, therefore, not directly explain a measurement of $\beta_\nu > 0$ for the lowest frequency LFI bands. There have been measured correlations between polarized synchrotron and dust emission \citep{Choi:2015xha} which could create intrinsic foreground $EB$ cross power spectra between synchrotron-dominated and dust-dominated channels. This could potentially bias our measurement of $\beta_\nu$ for frequencies where synchrotron emission dominates.

We used a filamentary dust model for the intrinsic $EB$ of polarized dust emission, but to what extent can we trust this model to describe the $EB$ of dust emission, especially at nearly full-sky? And does synchrotron emission have an intrinsic $EB$ power spectrum? If the filamentary dust ansatz describes dust sufficiently and synchrotron has a negligible $EB$, the measurements presented in this work provide evidence for a mostly frequency-independent cosmic birefringence angle, but we cannot claim that we have a good enough understanding of the polarized foreground emission yet. Hence, we do not claim any statistical significance with regard to our cosmic birefringence measurements.

More work needs to be done to understand the physics of the polarized foreground emission, but strong priors on the miscalibration angles of future satellite missions can also help us ascertain whether we are measuring the foreground or a cosmic signal.

\begin{acknowledgements}
    We are immensely grateful to Eiichiro Komatsu for guidance, suggestions, and discussions. We thank Hans Kristian Eriksen, Ingunn Wehus, and Duncan Watts for their comments on earlier drafts. We also thank Patricia Diego-Palazuelos and Yuto Minami for useful discussions and feedback on an earlier draft. We are also grateful to Matthieu Tristram and Raelyn M. Sullivan for useful discussions. \planck\ is a project of the European Space Agency (ESA) with instruments provided by two scientific consortia funded by ESA member states and led by Principal Investigators from France and Italy, telescope reflectors provided through a collaboration between ESA and a scientific consortium led and funded by Denmark, and additional contributions from NASA (USA). We acknowledge funding from the European Research Council (ERC) under the Horizon 2020 Research and Innovation Programme (Grant agreement No.~819478). Softwares: PolSpice \citep{Chon:2003gx}, healpy \citep{Gorski:2004by, Zonca2019}, Matplotlib \citep{Hunter:2007}, NumPy \citep{2020NumPy-Array}, CAMB \citep{Lewis:2000}, emcee \citep{ForemanMackey:2012ig}.
\end{acknowledgements}

\bibliographystyle{aa}
\bibliography{references}

\begin{thebibliography}{42}
\expandafter\ifx\csname natexlab\endcsname\relax\def\natexlab#1{#1}\fi

\bibitem[{Arvanitaki {et~al.}(2010)Arvanitaki, Dimopoulos, Dubovsky, Kaloper,
  \& March-Russell}]{Arvanitaki:2009fg}
Arvanitaki, A., Dimopoulos, S., Dubovsky, S., Kaloper, N., \& March-Russell, J.
  2010, Phys. Rev. D, 81, 123530

\bibitem[{{BeyondPlanck Collaboration} {et~al.}(2020){BeyondPlanck
  Collaboration}, {Andersen}, {Aurlien}, {Banerji}, {Bersanelli}, {Bertocco},
  {Brilenkov}, {Carbone}, {Colombo}, {Eriksen}, {Eskilt}, {Foss},
  {Franceschet}, {Fuskeland}, {Galeotta}, {Galloway}, {Gerakakis},
  {Gjerl{\o}w}, {Hensley}, {Herman}, {Iacobellis}, {Ieronymaki}, {Ihle},
  {Jewell}, {Karakci}, {Keih{\"a}nen}, {Keskitalo}, {Maggio}, {Maino}, {Maris},
  {Mennella}, {Paradiso}, {Partridge}, {Reinecke}, {San}, {Suur-Uski},
  {Svalheim}, {Tavagnacco}, {Thommesen}, {Watts}, {Wehus}, \&
  {Zacchei}}]{2020arXiv201105609B}
{BeyondPlanck Collaboration}, {Andersen}, K.~J., {Aurlien}, R., {et~al.} 2020,
  arXiv e-prints, arXiv:2011.05609

\bibitem[{Carroll {et~al.}(1990)Carroll, Field, \& Jackiw}]{Carroll:1989vb}
Carroll, S.~M., Field, G.~B., \& Jackiw, R. 1990, Phys. Rev. D, 41, 1231

\bibitem[{Challinor \& Chon(2005)}]{Challinor:2004pr}
Challinor, A. \& Chon, G. 2005, Mon. Not. Roy. Astron. Soc., 360, 509

\bibitem[{Choi \& Page(2015)}]{Choi:2015xha}
Choi, S.~K. \& Page, L.~A. 2015, JCAP, 12, 020

\bibitem[{Chon {et~al.}(2004)Chon, Challinor, Prunet, Hivon, \&
  Szapudi}]{Chon:2003gx}
Chon, G., Challinor, A., Prunet, S., Hivon, E., \& Szapudi, I. 2004, Mon. Not.
  Roy. Astron. Soc., 350, 914

\bibitem[{Clark {et~al.}(2021)Clark, Kim, Hill, \& Hensley}]{Clark:2021kze}
Clark, S.~E., Kim, C.-G., Hill, J.~C., \& Hensley, B.~S. 2021, Astrophys. J.,
  919, 53

\bibitem[{{Diego-Palazuelos} {et~al.}(2022){Diego-Palazuelos}, {Eskilt},
  {Minami}, {Tristram}, {Sullivan}, {Banday}, {Barreiro}, {Eriksen},
  {G{\'o}rski}, {Keskitalo}, {Komatsu}, {Mart{\'\i}nez-Gonz{\'a}lez}, {Scott},
  {Vielva}, \& {Wehus}}]{2022arXiv220107682D}
{Diego-Palazuelos}, P., {Eskilt}, J.~R., {Minami}, Y., {et~al.} 2022, \prl,
  128, 091302

\bibitem[{Foreman-Mackey {et~al.}(2013)Foreman-Mackey, Hogg, Lang, \&
  Goodman}]{ForemanMackey:2012ig}
Foreman-Mackey, D., Hogg, D.~W., Lang, D., \& Goodman, J. 2013, Publ. Astron.
  Soc. Pac., 125, 306

\bibitem[{Fujita {et~al.}(2021)Fujita, Murai, Nakatsuka, \&
  Tsujikawa}]{Fujita:2020ecn}
Fujita, T., Murai, K., Nakatsuka, H., \& Tsujikawa, S. 2021, Phys. Rev. D, 103,
  043509

\bibitem[{Galaverni {et~al.}(2015)Galaverni, Gubitosi, Paci, \&
  Finelli}]{Galaverni:2014gca}
Galaverni, M., Gubitosi, G., Paci, F., \& Finelli, F. 2015, JCAP, 08, 031

\bibitem[{Gorski {et~al.}(2005)Gorski, Hivon, Banday, Wandelt, Hansen,
  Reinecke, \& Bartelman}]{Gorski:2004by}
Gorski, K.~M., Hivon, E., Banday, A.~J., {et~al.} 2005, Astrophys. J., 622, 759

\bibitem[{Gubitosi {et~al.}(2014)Gubitosi, Martinelli, \&
  Pagano}]{Gubitosi:2014cua}
Gubitosi, G., Martinelli, M., \& Pagano, L. 2014, J. Cosmol. Astropart. Phys.,
  12, 020

\bibitem[{{Gubitosi} \& {Paci}(2013)}]{2013JCAP...02..020G}
{Gubitosi}, G. \& {Paci}, F. 2013, \jcap, 2013, 020

\bibitem[{Harari \& Sikivie(1992)}]{Harari:1992ea}
Harari, D. \& Sikivie, P. 1992, Phys. Lett. B, 289, 67

\bibitem[{Harris {et~al.}(2020)Harris, Millman, van~der Walt, Gommers,
  Virtanen, Cournapeau, Wieser, Taylor, Berg, Smith, Kern, Picus, Hoyer, van
  Kerkwijk, Brett, Haldane, Fernández~del Río, Wiebe, Peterson,
  Gérard-Marchant, Sheppard, Reddy, Weckesser, Abbasi, Gohlke, \&
  Oliphant}]{2020NumPy-Array}
Harris, C.~R., Millman, K.~J., van~der Walt, S.~J., {et~al.} 2020, Nature, 585,
  357–362

\bibitem[{Hivon {et~al.}(2002)Hivon, Gorski, Netterfield, Crill, Prunet, \&
  Hansen}]{Hivon:2002}
Hivon, E., Gorski, K.~M., Netterfield, C.~B., {et~al.} 2002, The Astrophysical
  Journal, 567, 2

\bibitem[{Huffenberger {et~al.}(2020)Huffenberger, Rotti, \&
  Collins}]{Huffenberger:2019mjx}
Huffenberger, K.~M., Rotti, A., \& Collins, D.~C. 2020, Astrophys. J., 899, 31

\bibitem[{Hunter(2007)}]{Hunter:2007}
Hunter, J.~D. 2007, Computing in Science \& Engineering, 9, 90

\bibitem[{Jones {et~al.}(2018)}]{C-BASS:2018dwc}
Jones, M.~E. {et~al.} 2018, Mon. Not. Roy. Astron. Soc., 480, 3224

\bibitem[{Kahniashvili {et~al.}(2008)Kahniashvili, Durrer, \&
  Maravin}]{Kahniashvili:2008va}
Kahniashvili, T., Durrer, R., \& Maravin, Y. 2008, Phys. Rev. D, 78, 123009

\bibitem[{Kamionkowski {et~al.}(1997)Kamionkowski, Kosowsky, \&
  Stebbins}]{Kamionkowski:1996zd}
Kamionkowski, M., Kosowsky, A., \& Stebbins, A. 1997, Phys. Rev. Lett., 78,
  2058

\bibitem[{{Lewis} {et~al.}(2000){Lewis}, {Challinor}, \&
  {Lasenby}}]{Lewis:2000}
{Lewis}, A., {Challinor}, A., \& {Lasenby}, A. 2000, Astrophys. J., 538, 473

\bibitem[{Marsh(2016)}]{Marsh:2015xka}
Marsh, D. J.~E. 2016, Phys. Rept., 643, 1

\bibitem[{Martire {et~al.}(2022)Martire, Barreiro, \&
  Mart\'\i{}nez-Gonz\'alez}]{Martire:2021gbc}
Martire, F.~A., Barreiro, R.~B., \& Mart\'\i{}nez-Gonz\'alez, E. 2022, JCAP,
  04, 003

\bibitem[{Minami \& Komatsu(2020{\natexlab{a}})}]{Minami:2020odp}
Minami, Y. \& Komatsu, E. 2020{\natexlab{a}}, Phys. Rev. Lett., 125, 221301

\bibitem[{Minami \& Komatsu(2020{\natexlab{b}})}]{MinamiKomatsu:2020}
Minami, Y. \& Komatsu, E. 2020{\natexlab{b}}, PTEP, 2020, 103E02

\bibitem[{Minami {et~al.}(2019)Minami, Ochi, Ichiki, Katayama, Komatsu, \&
  Matsumura}]{Minami:2019ruj}
Minami, Y., Ochi, H., Ichiki, K., {et~al.} 2019, PTEP, 2019, 083E02

\bibitem[{Myers \& Pospelov(2003)}]{Myers:2003fd}
Myers, R.~C. \& Pospelov, M. 2003, Phys. Rev. Lett., 90, 211601

\bibitem[{{Planck Collaboration Int. LVII}(2020)}]{Akrami:2020bpw}
{Planck Collaboration Int. LVII}. 2020, Astron. Astrophys., 643, A42

\bibitem[{{Planck Collaboration Int. XLIX}(2016)}]{Aghanim:2016fhp}
{Planck Collaboration Int. XLIX}. 2016, Astron. Astrophys., 596, A110

\bibitem[{{Planck Collaboration Int. XXII}(2015)}]{PlanckForeground:2014zja}
{Planck Collaboration Int. XXII}. 2015, Astron. Astrophys., 576, A107

\bibitem[{{Planck Collaboration IV}(2020)}]{PlanckComponentSeparation:2018}
{Planck Collaboration IV}. 2020, Astron. Astrophys., 641, A4

\bibitem[{{Planck Collaboration VI}(2020)}]{planckcosmo:2018}
{Planck Collaboration VI}. 2020, Astron. Astrophys., 641, A6, [Erratum:
  Astron.Astrophys. 652, C4 (2021)]

\bibitem[{{Planck Collaboration XI}(2020)}]{planckdust:2018}
{Planck Collaboration XI}. 2020, Astron. Astrophys., 641, A11

\bibitem[{Poidevin {et~al.}(2018)}]{QUIJOTE:2018ntj}
Poidevin, F. {et~al.} 2018, in {13th Rencontres du Vietnam}: {Cosmology 2017}

\bibitem[{Seljak \& Zaldarriaga(1997)}]{Seljak:1996gy}
Seljak, U. \& Zaldarriaga, M. 1997, Phys. Rev. Lett., 78, 2054

\bibitem[{Shore(2005)}]{Shore:2004sh}
Shore, G.~M. 2005, Nucl. Phys. B, 717, 86

\bibitem[{Subramanian(2016)}]{Subramanian:2015lua}
Subramanian, K. 2016, Rept. Prog. Phys., 79, 076901

\bibitem[{Thorne {et~al.}(2018)Thorne, Fujita, Hazumi, Katayama, Komatsu, \&
  Shiraishi}]{Thorne:2017jft}
Thorne, B., Fujita, T., Hazumi, M., {et~al.} 2018, Phys. Rev., D97, 043506

\bibitem[{Turner \& Widrow(1988)}]{Turner:1987bw}
Turner, M.~S. \& Widrow, L.~M. 1988, Phys. Rev. D, 37, 2743

\bibitem[{Zonca {et~al.}(2019)Zonca, Singer, Lenz, Reinecke, Rosset, Hivon, \&
  Gorski}]{Zonca2019}
Zonca, A., Singer, L.~P., Lenz, D., {et~al.} 2019, Journal of Open Source
  Software, 4, 1298

\end{thebibliography}

\end{document}